\documentclass[12pt,draftclsnofoot, onecolumn]{IEEEtran} 

\usepackage{amssymb}
\usepackage{amsthm}
\usepackage{graphicx}
\usepackage{array,color}
\usepackage{amsmath}
\usepackage{stfloats}
\usepackage{graphicx}
\usepackage{subfigure}
\usepackage{tabularx}
\usepackage{epsfig,epsf,color,balance,cite}
\usepackage{algorithmic}
\usepackage{algorithm}
\usepackage{bm}
\usepackage{textcomp}
\usepackage{color}
\usepackage{multirow}
\usepackage{cite}
\usepackage{enumerate}
\usepackage{cases}
\usepackage{color}
\usepackage{url}
\usepackage{epstopdf}
\usepackage{ textcomp}

\begin{document}
	\title{Self-Sustainable Reconfigurable Intelligent Surface Aided Simultaneous Terahertz Information and Power Transfer (STIPT)}
	\author{\IEEEauthorblockN{Yijin Pan, Kezhi Wang, Cunhua Pan, Huiling Zhu and Jiangzhou Wang,~\IEEEmembership{Fellow,~IEEE}}
	\thanks{Y. Pan is with the National Mobile Communications Research Laboratory, Southeast University, Nanjing 211111, China, and also with School of Engineering and Digital Arts, University of Kent, UK.}
	\thanks{K. Wang is with the Department of Computer and Information Sciences, Northumbria University, UK.}
	\thanks{C. Pan is with the School of Electronic Engineering and Computer Science, Queen Mary, University of London, UK.}
	\thanks{J. Wang and H. Zhu are with the School of Engineering and Digital Arts, University of Kent, UK.}}
	\maketitle

\begin{abstract}
This paper proposes a new simultaneous terahertz (THz) information and power transfer (STIPT) system, which is assisted by reconfigurable intelligent surface (RIS) for both the information data and power transmission.
We aim to maximize the information users' (IUs') sum data rate while guaranteeing the energy users' (EUs') and RIS's power harvesting requirements.
To solve the formulated non-convex problem, the block coordinate descent (BCD) based algorithm is adopted to alternately optimize the transmit precoding of IUs, RIS's reflecting coefficients, and RIS's coordinate.
The Penalty Constrained Convex Approximation (PCCA) Algorithm is proposed to solve the intractable optimization problem of the RIS's coordinate, where the solution's feasibility is guaranteed by the introduced penalties.
Simulation results confirm that the proposed BCD algorithm can significantly enhance the performance of STIPT by employing RIS.
\end{abstract}


\newpage

\section{Introduction}

The current mobile network is experiencing an unprecedented evolution with the increasing number of attractive mobile applications, which results in expectations for extremely high data rates for realizing a variety of multimedia services\cite{You.2021}.
Thanks to the abundant spectrum resources in the terahertz (THz) band, the THz communication is able to realize high transmission rates from hundreds of Gbps to several Tbps.
As a result, THz transmission is envisioned as an emerging solution to meet the ultra-high-speed data rate demands of the enhanced mobile broadband (eMBB) services, such as virtual reality and high-definition data streams\cite{Barros.2017}.

At the same time, the wireless power harvesting (WPH) technology provides an attractive approach for the Internet of things (IoT) devices with limited battery life to reap power from external radio frequency sources \cite{Zhang.2013}.
The WPH technique promises the potentials for replacing batteries in low power consumption devices or increasing their battery lifespans \cite{Zhang.2019}.
With these properties, numerous devices and sensors are being manufactured with the WPH power supply mode, including the nanoscale IoT devices\cite{Akhtar.2020}.
Nevertheless, the traditional low-frequency radio waves may no longer be suitable for WPH \cite{Mizojiri.2018b}.
This is because the low-frequency radio wave normally requires a large antenna aperture to capture a large portion of radiated electromagnetic wave due to the relatively large wavelength\cite{Tran.2017}, which tends to exceed the size limit of the IoT devices and sensors \cite{Rong.2017},  especially for the nanoscale IoT devices.
To tackle this issue, a possible solution is to increase the frequency of the radio so that the size of antennas can be miniaturized and the transmission beam directivity can be improved\cite{Mizojiri.2018b}.

Fortunately, as a key component of the 6G communication system, the THz band bridges the gap between mmWave and optical band.
Specifically, THz band ranges from 100 GHz to 10 THz such that the wavelength of THz can greatly reduce the required antenna aperture \cite{JingboTan}.
This makes utilizing the THz electronics for wireless power harvesting a very attractive approach. 
Also, emerging rectennas have been proposed and manufactured which make the energy harvesting in THz bands becomes possible\cite{Mizojiri.2018,Mizojiri.2019}.
As a result, combining the benefits of the THz transmission in providing high-speed data rates with WPH for IoT devices will be a promising research direction, leading to a new term of Simultaneous Terahertz Information and Power Transfer (STIPT) network.
 
\subsection{Related Works}

In the THz communication system, due to the ultra-high radio frequency, the THz links are easily blocked by obstacles in transmission paths. 
This feature of the THz band greatly limits the transmission distance, which necessitate establishing efficient STIPT communications.
To address this issue, it has been proposed to utilize the reconfigurable intelligent surface (RIS), also known as intelligent reflective surface (IRS) to help compensate for the blocked communication links\cite{Wu.202076}.
By adjusting the reflecting coefficients of RIS, the propagation channel condition can be significantly improved to enhance the system performance\cite{DiRenzo.2020}.

Recently, extensive efforts have been devoted to numerous applications of the RIS-assisted transmissions\cite{Basar.2019}.
A comprehensive survey about these applications of the RIS has been given in \cite{Gong.2020}, 
and the potential benefits that can be brought by the RIS has been explained in various aspects in \cite{CunhuaPan.2020}.
In \cite{9090356}, the RIS has been exploited to enhance the cell-edge performance in multicell MIMO communication systems.
Also, the RIS has been utilized to enhance latency performance of the mobile edge computing (MEC) system in \cite{Bai.2020}.
Furthermore, the RIS was also utilized to enhance the physical layer security by improving the secrecy rate \cite{Cui.2019,Dong.2020,Feng.2020,Hong.2020,Yu.122019} and reducing the transmit power \cite{Chu.2020}.
The RIS-enhanced orthogonal frequency division multiplexing (OFDM) system and its corresponding  transmission protocol were investigated in \cite{Yang.2020b}.

As for the simultaneous wireless information and power transfer (SWIPT) system, there are already a few contributions on utilizing the RIS to enhance the performance of SWIPT systems\cite{Feng.202117}.
An RIS-aided MIMO broadcasting SWIPT system was investigated in \cite{pan2020intelligent}, where the transmit precoding matrices and passive phase shift matrix of the RIS were jointly optimized to maximize the weighted sum rate of information users while guaranteeing the users' energy harvesting requirement.
The contributions in \cite{Hehao.2020} and \cite{Liu.2020} investigated the RIS-aided secure transmission system, where the obtained secure rate and energy efficiency were enhanced by optimizing the reflecting coefficients of RIS, respectively.  
 In \cite{Li.20201130}, RIS was leveraged to enhance the performance of non-orthogonal multiple access (NOMA) and the wireless power transfer (WPT) efficiency of SWIPT.
The energy harvesting efficiency of the RIS-assisted MIMO broadcasting SWIPT system was investigated in \cite{Wu.2020}, where the total transmit power required at the AP was minimized while satisfying the QoS constraints of the information users and the energy users.
A RIS-assisted wireless power transfer OFDM-based MEC system was investigated in \cite{Bai.2020311}, where the total energy consumption was minimized by optimizing the power allocation and computation resource allocation.

It is worth pointing out that the above works only considered the micro/millimeter wave communications.
Compared with traditional micro/millimeter transmission system, the THz band transmissions suffer from high molecular absorption and propagation loss\cite{Jornet.2011}.
The extremely short wavelengths of THz signals makes the obstacles in the path tend to absorb THz signals rather than reflect them\cite{Pengnoo.2020}. 
Therefore, the RIS is envisioned to be a necessity for the future THz communications to bypass blockages\cite{Ma.20201213}.
Most recently, specific efforts have been devoted to the use of RIS to enhance the THz transmission performance\cite{Ning.2019d,Ma.2020672020611}.
For instance, the coverage analysis in \cite{Boulogeorgos.2021} has highlighted the impact of the molecular absorption loss on the path loss of the RIS-assisted THz channel.
As the THz band appears to be frequency-selective with many path loss peaks,
the sum rate of an RIS-assisted THz system was maximized in \cite{Pan.2020}, where a whole band is divided into several sub-bands.
The passive reflecting phase shifters of RIS was investigated in \cite{Ning.122019} to enhance the secrecy rate of THz communication.
An RIS-aided multi-user THz MIMO system with orthogonal frequency division multiple access was investigated in \cite{Hao.2020913}, where the weighted sum rate was maximized by jointly optimizing the hybrid beamforming and reflecting matrix of the RIS.
As the RIS is able to control the propagation direction of THz waves for mitigating the blockage issue, the RIS was utilized in \cite{Pan.20201027} to assist the UAV THz transmission, where the joint passive beamforming design and trajectory optimization were investigated.
In \cite{Pan.20201027}, as the available spectrum in the THz band varies with the link distance, the unique channel fading characteristics of the THz channel was exploited to optimize the UAV trajectory.

\subsection{Motivations and Contributions}

Most of the above-mentioned RIS-aided transmission approaches are not specifically designed for STIPT systems, as integrating RIS into the STIPT face many challenges.

First of all, in the THz band, many path loss peaks appear and therefore the total band is divided into many sub-bands \cite{Han.2016}.
According to \cite{Pan.2020}, the locations of these path loss peaks vary with the carrier frequencies and the transmission distance. 
Consequently, it becomes challenging to efficiently utilize the frequency-selective THz band, and the precoding for different users needs to be designed carefully on multiple sub-bands.
More importantly, the link distance is dependent on the location of the RIS \cite{Boulogeorgos.20186252018628}, but the impact of the RIS's location is generally ignored in the current RIS applications.
In a typical STIPT application scenario, the energy users are normally IoT devices or sensors, and the information users are deployed for the monitor tasks such as transmitting high-definition video/figures that entail ultra-high data rate. 
The locations of energy users and information users can be available at the network AP.
Consequently, in this scenario where the positions of users are relatively fixed, optimization of the RIS's location has the potential to effectively compensate for the link-distance dependent fading in the THz band.
In fact, optimizing the location of RIS with the power constraints is challenging due to multiple periodic cosine components in the channel expression, and currently, no efficient solutions have been reported in the literature, to the best of our knowledge. 

Secondly, due to the high radio frequencies in the THz band, the number of RIS elements can be significantly increased within a limited area in order to provide a better reflecting performance. 
According to \cite{Huang.2019}, the RIS power consumption depends on the type and the number of reflecting elements. 
Therefore, the energy consumption of RIS cannot be ignored in this case, and it is proportional to the number of RIS elements\cite{lyu2020optimized}.
In \cite{Hu.2021126b}, a part of the RIS's elements are selected to harvest the received energy and the remaining elements help the secure information transmission, but this approach was designed for the micro/millimeter transmissions.
However, in the current RIS-assisted SWIPT schemes, the power consumption of the RIS is generally ignored,  and it is also unclear how to power the RIS  in the THz transmission system.

As a result, it is imperative to jointly consider the impact of the RIS's power consumption, reflecting coefficients, and its location on the STIPT's system.
The above mentioned key fundamental issues need to be resolved, and how to obtain an efficient STIPT communication system is still unknown.


Against the above background, in this paper, we consider the downlink transmission of the STIPT network, where the information users (IUs) and energy users (EUs) are jointly served by the RIS-assisted THz links. 
In our system, the RIS is also equipped with the WPH module to harvest energy from the received THz radios to maintain its circuit power consumption.  
The precoding for IUs, the RIS's reflecting coefficients, and the RIS's coordinate are jointly optimized to maximize the IUs' achievable rates while satisfying the EU's and RIS's power harvesting requirements.
Overall, our contributions can be summarized as follows:
\begin{itemize}
 
\item We propose an RIS-aided STIPT system to simultaneously transmit information and power for IUs and EUs, respectively.
The RIS-assisted THz channel is modelled as a function of the RIS reflecting coefficients and the coordinate of RIS.
The RIS can harvest power from the received radio, where the harvested power can be adjusted by the amplitude of its reflecting coefficients.

\item The optimization problem is formulated to maximize the IUs' sum rates while guaranteeing the power harvesting requirements of the EUs.
The original non-convex problem is first reformulated by utilizing the equivalence between the weighted minimum mean-square error (WMMSE) and the signal-to-noise ratio (SINR).
Then, we decouple the optimization problem into three subproblems: optimization of the precoding for IUs, RIS's reflecting coefficients and RIS's coordinate.

\item The precoding for IUs and RIS's reflecting coefficients are obtained by utilizing the successive convex approximation method.
To deal with the intractable optimization problem of the RIS's coordinate, we propose the Penalty Constrained Convex Approximation (PCCA) Algorithm to guarantee the solution's feasibility and the convergence of the block coordinate descent (BCD) algorithm.

\item  Extensive simulation results are provided to show the performance gain achieved by the proposed STIPT system compared with benchmarks.
It is shown that the sum rate performance of IUs is greatly affected by the RIS's coordinate.
By utilizing the proposed BCD algorithm, the THz channel can be optimized to fully exploit the spatial diversity so that the sum rate performance can be significantly enhanced. 
\end{itemize}

The reminder of this paper is organized as follows:
Section II describes the system model of the proposed RIS-aided STIPT system and formulates the optimization problem.
Sections III develops the detailed algorithm to solve the formulated sum-rate maximization problem.
In Section IV, the simulation results are presented to show the performance gain and the impact of system parameters, and Section V concludes the paper.

\textit{Notation}: For a vector $\bm{x}$, $|\bm{x}|$ and $(\bm{x})^T$ respectively denote its Euclidean norm and its transpose.
$c$ represents the light speed.
For matrix $\bm{A}$, $\bm{A}^*$ and $\bm{A}^{\star}$ represent the conjugate operator and converged solution, respectively.
$\mathbb{C}^{M \times 1}$ denotes the set of $M \times 1$ complex vectors. 
$\text{diag}(\bm{X})$ represents the vector that is obtained from the diagonal entries of matrix $\bm{X}$.
$\bm{a} \odot \bm{b} $ represents the Hadamard (point-wise) product of $\bm{a}$ and $\bm{b}$.

\begin{figure}
	\centering
	\vspace{-1em}
	\includegraphics[width=0.6\textwidth]{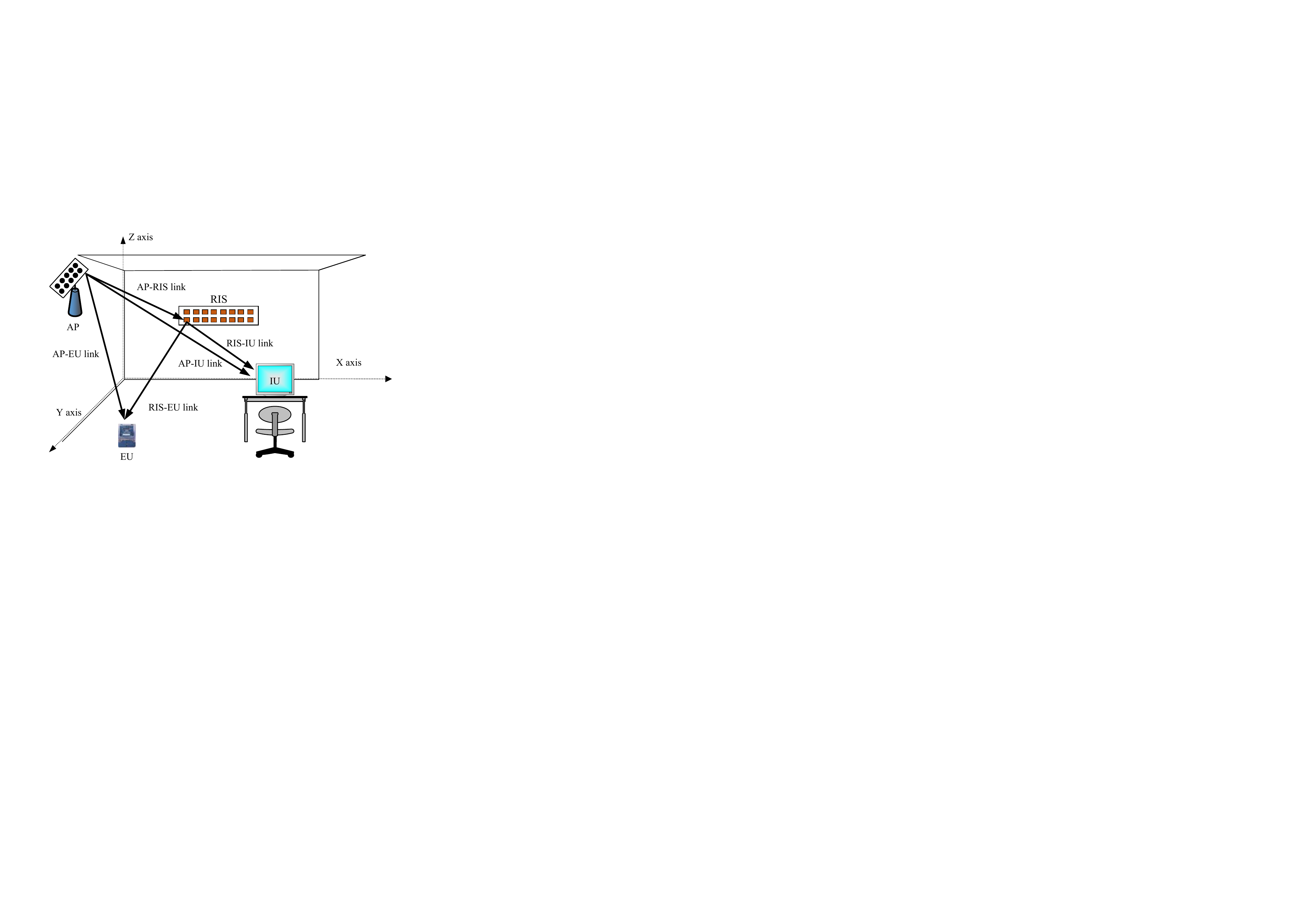}
	\vspace{-1em}
	\caption{The RIS-assisted SWIFT system.}
	\vspace{-1em}
	\label{fig1}
\end{figure}

\section{System Model and Problem Formulation}

Consider the downlink of a STIPT system that needs to serve the IoT sensors and video transmission devices at the same time, as shown in Fig. \ref{fig1}.
The network access point  (AP) operates in the THz band. 
The IoT sensors are EUs and they need to harvest power from the THz radio, meanwhile, IUs require high-speed data transmission for transmitting high-definition video/figures.

The AP is equipped with $N_t$ transmit antennas to serve $I$ IUs and $M$ EUs, and all the IUs and EUs are equipped with $N_r$ receive antennas.
Let $\mathcal{I}$ and $\mathcal{M}$ respectively represent the set of IUs and EUs.
Then, the set of total users is given by $\mathcal{U} = \mathcal{I} \cup \mathcal{M}$, and the total number of all users (including EU and IU) is denoted by $U = \left| {\cal U} \right|$.
In the following, the user $u$ can be either EU or IU.

As the THz channel is frequency-selective, the total THz band is divided into $K$ sub-bands (SBs).
Let $f_k$ denote the central frequency of SB $k$, and its wavelength is given by $\lambda_k = \frac{c}{f_k}$.

Normally, the wireless transmission channel includes the line-of-sight (LOS) link and non-line-of-sight (NLOS) links, where NLOS links consist of reflected, scattered, and diffracted components. 
As the scattered and diffracted components are shown to play insignificant roles in the received signal power in \cite{Han.2016,Priebe.2013}, similar to \cite{Du.2020,Hao.2020913}, the scattered and diffracted rays are neglected in the channel model.
In addition, according to \cite{Pengnoo.2020}, the surfaces of walls and ceilings appear ‘‘rough’’ for the THz signals so that they tend to absorb and scatter the THz signals rather than reflect them.
As the RIS is specially designed to enable redirecting the incoming signal to the desired directions, the NLOS components in this work are only contributed by reflected paths from the RIS.

\begin{figure}
	\subfigure[LOS links from AP's transmit antenna elements to user $u$'s first antenna element]
	{\begin{minipage}[t]{0.45\textwidth}
			\centering
			\includegraphics[width=1\textwidth]{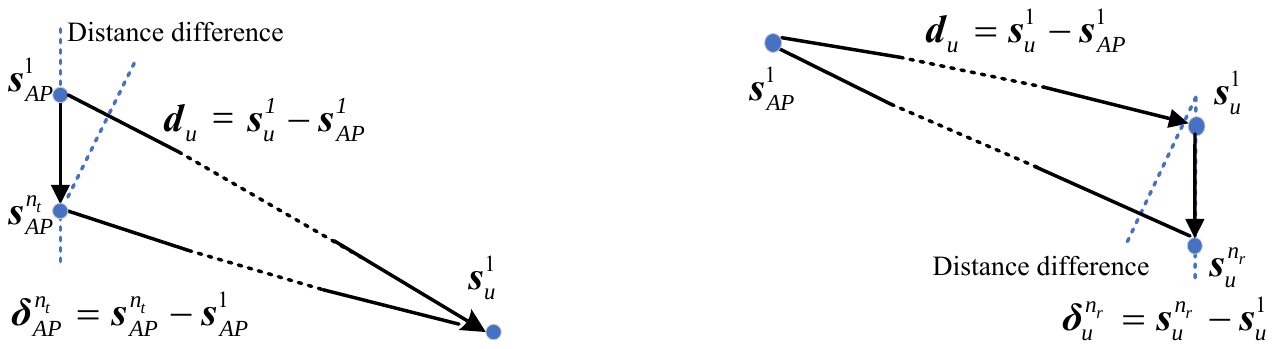}
	\end{minipage}}
	\subfigure[LOS links between user $u$'s receive antenna elements and AP's first antenna element]
	{\begin{minipage}[t]{0.45\textwidth}
			\centering
			\includegraphics[width=1\textwidth]{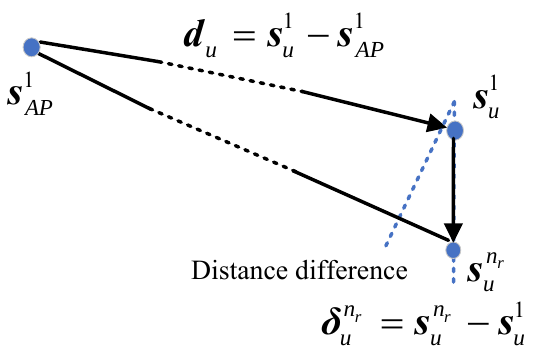}
	\end{minipage}}
	\centering
	\caption{The LOS links between AP and user $u$. } \label{FigDif}
\end{figure} 

\subsection{LOS Links without RIS}

Fig. \ref{FigDif} shows the LOS links between the AP and user $u$.
As shown in Fig. \ref{FigDif}, the coordinate of AP's first antenna element is denoted as $\bm{s}^1_{AP}\in \mathbb{R}^{3 \times 1}$, and the coordinate of user $u$'s first antenna element is denoted as $\bm{s}^1_{u}\in \mathbb{R}^{3 \times 1}$.
Then, the distance between the AP and the users is calculated as $|\bm{d}_u|$, where $\bm{d}_u = \bm{s}^1_{u} - \bm{s}^1_{AP}$ is the transmit vector for user $u$.
In addition, as shown in Fig. \ref{FigDif}, the transmission distances between different transmit (receive) antenna elements are different, which causes the phase difference between the channel gains in the frequency domain.
These phase differences are featured by the transmit array vector of the AP and the receive antenna vector of the user.

Fig. \ref{FigDif} (a) shows the distance differences between the transmit antennas of the AP.
In Fig. \ref{FigDif} (a), $\bm{s}^{n_t}_{AP} \in \mathbb{R}^{3 \times 1}$ denotes the coordinate of AP's $n_t$-th antenna element.
We define directional vector $\bm{\delta}^{n_t}_{AP} = \bm{s}^{n_t}_{AP} - \bm{s}^1_{AP} , n_t = 1, \cdots, N_t$, and $\bm{\delta}^{1}_{AP} = [0,0,0]^T$.
As shown in Fig. \ref{FigDif} (a), the phase difference $\theta_{k,u}^{n_t}$ between AP's $n_t$-th antenna element and the first element is evaluated as
\begin{equation}\label{theta}
\theta_{k,u}^{n_t} = \frac{2\pi f_k}{c} \frac{(\bm{d}_u)^T \bm{\delta}^{n_t}_{AP}}{|\bm{d}_u|}=\frac{2\pi}{\lambda_k} \frac{(\bm{d}_u)^T \bm{\delta}^{n_t}_{AP}}{|\bm{d}_u|}, n_t = 1, \cdots, N_t.
\end{equation}
Then, the transmit array vector from the AP to user $u$ on SB $k$ is then denoted by 
\begin{equation}
\bm{v}_{k,u} = [1, \cdots, \exp(-j\theta_{k,u}^{n_t}), \cdots,\exp(-j\theta_{k,u}^{N_t})]^T.
\end{equation}

Fig. \ref{FigDif} (b) shows the distance differences between the receive antennas of the user $u$.
In Fig. \ref{FigDif} (b), $\bm{s}_u^{n_r}\in \mathbb{R}^{3 \times 1}$ denote the coordinate of the $n_r$-th antenna element of user $u$.
Define the directional vector $\bm{\delta}^{n_r}_{u} =\bm{s}_u^{n_r} - \bm{s}_u^1, n_r = 1, \cdots, N_r,$ for the $n_r$-th receive antenna element.
As shown in Fig. \ref{FigDif} (b), the phase difference $\varrho^{n_r}_{k,u}$ between user's $n_r$-th antenna element and the first element is evaluated as
\begin{equation} \label{rho}
\varrho^{n_r}_{k,u}= \frac{2\pi f_k}{c} \frac{(\bm{d}_u)^T \bm{\delta}^{n_r}_{u}}{|\bm{d}_u|} = \frac{2\pi}{\lambda_k} \frac{(\bm{d}_u)^T \bm{\delta}^{n_r}_{u}}{|\bm{d}_u|}, n_r = 1, \cdots, N_r.
\end{equation}
Then, the receive array vector from user $u$ to the AP on SB $k$ is  
\begin{equation}
\bm{r}^{dir}_{k,u} = [1, \cdots, \exp(-j\varrho^{n_r}_{k,u}), \cdots,\exp(-j\varrho^{N_r}_{k,u})]^T.
\end{equation}

According to the ray tracing techniques \cite{Han.2015}, the path gain from the AP to user $u$ on SB $k$ is evaluated as 
\begin{equation}
{h}_{k,u} = \left(\frac{G_rG_t\lambda_k}{4 \pi |\bm{d}_u|} \right) \exp\left({-j 2\pi \frac{|\bm{d}_u|}{\lambda_k}}\right) \exp\left({-\frac{1}{2}K(f_k)|\bm{d}_u|}\right),
\end{equation}
where $G_r$ and $G_t$ respectively represents the antenna gain of transmit array and the antenna gain of receiving array, $|\bm{d}_u|$ is the distance from IU $u$ to the AP, and $K(f_k)$ is the overall absorption coefficient of the transmission medium on SB $k$.
Then, the LOS channel from the AP to user $u$ on SB $k$ is denoted by 
\begin{equation}\label{hku}
\bm{H}_{k,u} =  {h}_{k,u}\bm{r}^{dir}_{k,u}\bm{v}^H_{k,u}, u \in \mathcal{I}\cup \mathcal{M}. 
\end{equation}

\subsection{RIS Assisted NLOS Links}\label{RISH}

The number of reflecting elements of the RIS is $N$. 
The coordinate of the RIS's first reflecting element is denoted as $\bm{s}_{RIS}^1\in \mathbb{R}^{3 \times 1}$.

We first consider the link from the AP to the RIS.
The transmit distance between the AP and the RIS is evaluated as $|\bm{d}_0|$, where the transmit vector is $\bm{d}_0 =\bm{s}_{RIS}^1 - \bm{s}_{AP}^1$.
Similar to (\ref{theta}), the phase difference $\theta_{k}^{n_t}$ between AP's $n_t$-th antenna element and the first element is evaluated as
\begin{equation}
\theta_{k}^{n_t} = \frac{2\pi}{\lambda_k} \frac{(\bm{d}_0)^T \bm{\delta}^{n_t}_{AP}}{|\bm{d}_0|}, n_t = 1, \cdots, N_t. 
\end{equation}
Then, the transmit array vector from the AP to the RIS on SB $k$ is then denoted by 
\begin{equation}
\bm{v}_{k} = [1, \cdots, \exp(-j\theta_{k}^{n_t}), \cdots,\exp(-j\theta_{k}^{N_t})]^T.
\end{equation}

In addition, due to different transmit distance between the reflecting elements, we define the  receiving array vector $\bm{e}_{k}$ to feature the 
relative phase differences between the signals received on SB $k$ at different reflecting elements.

Similar to (\ref{rho}), the directional vector is defined as $\bm{\delta}^{n}_{RIS} =\bm{s}_{RIS}^n - \bm{s}_{RIS}^1$, $n = 1, \cdots, N$,
 and the phase difference between RIS's $n$-th reflecting element and the first reflecting element is
\begin{equation}
\vartheta_{k}^{n}  = \frac{2\pi}{\lambda_k} \frac{(\bm{d}_0)^T \bm{\delta}^{n}_{RIS} }{|\bm{d}_0|},  n = 1, \cdots, N. 
\end{equation}

Then, the receive array vector is given by 
\begin{equation}
\bm{e}_{k} = \left [ 1, \cdots, \exp(-j\vartheta_{k}^{n}), \cdots,\exp(-j\vartheta_{k}^{N})\right ]^T.
\end{equation}
As the path-loss gain from  the AP to RIS on SB $k$  is evaluated as
\begin{equation}
H_k = \left( \frac{G_t\lambda_k}{4 \pi |\bm{d}_0|} \right) \exp\left({-j 2\pi \frac{|\bm{d}_0|}{\lambda_k}}\right) \exp\left({-\frac{1}{2}K(f_k)|\bm{d}_0|}\right), \label{Hk}
\end{equation}
then the LOS channel from the AP to RIS on SB $k$ is denoted by 
\begin{equation}\label{HRIS}
\bm{H}_{k} = H_k\bm{e}_{k}\bm{v}_{k}^H. 
\end{equation}

Then, we consider the links from the RIS to the users.
The transmit distance between the RIS and the user $u$ is $|\bm{d}_{0,u}|$, where $\bm{d}_{0,u} =  \bm{s}_u^1 - \bm{s}_{RIS}^1$.
Similarly, the phase difference between the $n$-th reflecting element and the first reflecting element is 
\begin{equation}
\vartheta^n_{k,u} = \frac{2\pi}{\lambda_k} \frac{(\bm{d}_{0,u})^T \bm{\delta}^{n}_{RIS} }{|\bm{d}_{0,u}|},  n = 1, \cdots, N,
\end{equation}
and the transmit array vector from the RIS to user $u$ on SB $k$ is expressed as
\begin{equation}
\bm{e}_{k,u} = \left [1, \cdots,\exp(-j\vartheta^n_{k,u}), \cdots,\exp(-j\vartheta^N_{k,u}) \right ]^T.
\end{equation}

As the power consumption of the RIS cannot be ignored, we assume that the WPH module is equipped in the RIS so that the RIS can also harvest energy from the radios sent by AP.
As a result,  the reflecting coefficient is denoted by $\Phi_n =\beta_n\exp(j\phi_{n})$, where $\beta_n$ and $\phi_{n}$ respectively represent the amplitude and the phase shift of the $n$-th reflecting element.
Then, the phase shift matrix of the RIS is denoted by 
\begin{equation}
\bm \Phi= \text{diag} \{\beta_n\exp(j\phi_{n}),n = 1, \cdots, N\} .
\end{equation}
We then have the following constraints for the reflecting coefficients as
\begin{equation}
C1 :|\beta_n\exp(j\phi_{n})|  \leq 1, n= 1,\cdots, N.
\end{equation}

In addition, the phase difference at SB $k$ between user $u$'s first receive antenna element and the $n_r$-th element is 
\begin{equation}
\zeta^{n_r}_{k,u} = \frac{2\pi}{\lambda_k} \frac{(\bm{d}_{0,u})^T \bm{\delta}^{n_r}_{u}}{|\bm{d}_{0,u}|} , n_r = 1, \cdots, N_r.
\end{equation}
Then, the receive array vector from user $u$ to RIS on SB $k$ is expressed as 
\begin{equation}
\bm{r}_{k,u} = \left [1, \cdots,\exp(-j\zeta^{n_r}_{k,u}), \cdots, \exp(-j\zeta^{N_r}_{k,u})\right ]^T.
\end{equation}

The cascaded channel gain of the AP-RIS-user $u$ link on SB $k$ can be expressed as \cite{Tang.2019}
\begin{equation}
{g}_{k,u}= \left( G_tG_r\frac{\lambda_k}{8 \sqrt{\pi^3} |\bm{d}_{0,u}| |\bm{d}_0|} \right) \exp\left({-j 2\pi \frac{|\bm{d}_{0,u}| + |\bm{d}_0|}{\lambda_k}}\right) \exp\left({-\frac{1}{2}K(f_k)(|\bm{d}_{0,u}| + |\bm{d}_0|)}\right), \label{Gku}
\end{equation}
where $|\bm{d}_{0,u}|$ and $|\bm{d}_0|$ represent the distance from the RIS to the user $u$ and the AP, respectively. 
Overall, the AP-RIS-user $u$ link on SB $k$ is given by
\begin{equation}\label{gku}
\bm{G}_{k,u} = g_{k,u}\bm{r}_{k,u}\bm{e}_{k,u}^H \bm \Phi \bm{e}_{k}\bm{v}_{k}^H, u \in \mathcal{I}\cup \mathcal{M}.
\end{equation} 

\subsection{Information Transfer}

The signal vector transmitted from the AP to IU $i$ on SB $k$ is $\bm{s}_{k,i} \in \mathbb{C}^{d \times 1}$.
Suppose that the data symbol $\bm{s}_{k,i}$ satisfies $\mathbb{E}[\bm{s}_{k,i}\bm{s}^*_{k,i}]=\bm{I}_d$ and $\mathbb{E}[\bm{s}_{k,i}\bm{s}^*_{k,j}]=\bm{0}$ for $i \neq j$.
Let $\bm{F}_{k,i} \in \mathbb{C}^{N_t \times d}$ denote the precoding matrix used by the AP for IU $i$ on SB $k$. 
Then, the transmitted signal $\bm{x}_k  \in \mathbb{C}^{N_t \times 1} $ from the AP on SB $k$ is  
\begin{equation}
\bm{x}_k = \sum_{i =1}^{I} \bm{F}_{k,i} \bm{s}_{k,i}.
\end{equation}
With the aid of the RIS, the received signal at the IU $i$ on SB $k$ is 
\begin{equation}
\bm{y}_{k,i} = ( \bm{H}_{k,i} + \bm{G}_{k,i})\bm{x}_k  + \bm{n}_{k,i} =  \bm{Z}_{k,i}\bm{x}_k  + \bm{n}_{k,i}, 
\end{equation}
where $ \bm{Z}_{k,i} = \bm{H}_{k,i} + \bm{G}_{k,i}$, and $\bm{n}_{k,i}$ is the additive Gaussian noise.

Then, the achievable data rate of IU $i$ on SB $k$ is given by 
\begin{equation}
R_{k,i} = \log\left |\bm{I}_{N_r} + \bm{F}^H_{k,i}\bm{Z}^H_{k,i}\bm{Z}_{k,i}\bm{F}_{k,i}\bm{J}_{k,i}^{-1}\right|, 
\end{equation}
where $\bm{J}_{k,i} = \sum_{u \neq i}^I\bm{Z}_{k,i}\bm{F}_{k,u}\bm{F}^H_{k,u}\bm{Z}^H_{k,i} + \sigma_{k,i}^2\bm{I}_{N_r}$, and $\sigma_{k,i}^2$ is the noise power.

Furthermore, as the transmit power is limited, we have the following constraints for the precoding matrices:
\begin{equation}
C2 : \sum_{k=1}^K\sum_{i =1}^{I}\|\bm{F}_{k,i} \|^2_F \leq  P_T^{max}.
\end{equation}

\subsection{Energy Harvesting}

As the RIS also harvests energy from the AP, so that the reflecting coefficients can be adjusted to satisfy the energy harvesting requirement.
That is to say, a part of the AP’s energy is reflected by the RIS, and the remaining part is fed into the RIS’s WPH unit for harvesting.
The power received by RIS on SB $k$ is 
\begin{equation}
q^{in}_{k} = \sum_{i =1}^I \text{tr}\left(\bm{F}^H_{k,i}\bm{H}_{k}^H\bm{H}_{k}\bm{F}_{k,i}\right).
\end{equation}
The reflected power by RIS on SB $k$ is 
\begin{equation}
q^{out}_{k} = \sum_{i =1}^I \text{tr}\left(\bm{F}^H_{k,i}\bm{H}^H_{k}{\bm{\Phi}^H\bm{\Phi}}\bm{H}_{k}\bm{F}_{k,i}\right).
\end{equation}
Then, the harvested power by RIS is calculated by 
\begin{equation}
P_{RIS} = \sum_{k=1}^K \eta_k(q^{in}_{k}-q^{out}_{k}).
\end{equation}
where $\eta_k$ denotes the power harvesting efficiency on SB $k$, since the RF-DC conversion efficiency is dependent on the carrier's frequency.
Let $P^{I}$ denote the required power for RIS, then we have
\begin{equation}
C3 : \sum_{k=1}^K \sum_{i =1}^I\eta_k \text{tr}\left( \bm{F}^H_{k,i}\bm{H}^H_{k}(\bm{I}-\bm{\Phi}^H\bm{\Phi})\bm{H}_{k}\bm{F}_{k,i} \right)\geq P^{I}.
\end{equation}

Similarly, the power harvested by EU $m$ should satisfy the following constraint:
\begin{equation}
C4: \sum_{k=1}^{K} \sum_{i =1}^I\eta_k\text{tr}\left(\bm{Z}_{k,m}\bm{F}_{k,i}\bm{F}^H_{k,i}\bm{Z}^H_{k,m} \right)\geq P^{U}_m, m \in\mathcal{M},
\end{equation}
where $P^{U}_m$ is the required power of EU $m$, and $\bm{Z}_{k,m}$ is the composite channel gain between the AP and the EU $m$ on SB $k$.

\subsection{Problem Formulation}

For simplicity, we define $\bm{\beta} = [\beta_1, \cdots, \beta_N]$ as the amplitude vector of the reflecting coefficients, and define 
$\bm{\phi}= [\phi_{1}, \cdots, \phi_{N}]$ as the phase shifts of the reflecting coefficients.
For ease of presentation, in the following, we utilize the notation $\bm{L}$ to represent the coordinate of RIS,  which is optimized in the following section.
In the system model, we take the coordinate of RIS's first reflecting element, i.e., $ \bm{s}_{RIS}^1$, as the reference coordinate in the system model.
That is to say, $\bm{L}$ is equivalent to $ \bm{s}_{RIS}^1$ in the following.

It is observed that the sum rate of IUs and the harvested power for EUs are dependent on the coordinate of RIS $\bm{L}$, the transmit precoding matrices and the reflecting coefficients of the RIS.
Then, we can formulate the problem as:
\begin{equation}\label{Prom1}
	\begin{split}
	\underset{\bm{\beta}, \bm \phi,\bm{L},\bm{F}_{k,i} }{\text{max} }  
	&\quad \quad  R_s =\sum_{k=1}^K\sum_{i=1}^I R_{k,i}  \\
	\text{s.t.} & \quad  C1-C4. 
	\end{split}
\end{equation}
It is observed that Problem (\ref{Prom1}) is non-convex and difficult to solve due to the following reasons.
First of all, the optimization variables are coupled together and the objective function is intractable.
Moreover, according to the RIS-assisted channel model in Section \ref{RISH},  there is a complicated relationship between the position of RIS, the RIS's reflecting coefficients and the channel gain.
Therefore, effective reformulation and simplification are required to tackle the above optimization problem.

\section{Solution Analysis}

The original Problem (\ref{Prom1}) is non-convex and challenging to solve, we first reformulate the problem by leveraging the equivalence between the minimum mean-square error (MMSE) and the signal-to-noise ratio (SINR).
At IU $i$, the receive decoding matrix $\bm{U}_{k,i} \in \mathbb{C}^{N_r \times d }$ is applied to the  received signal on SB $k$ so that $\hat{\bm{s}}_{k,i} = \bm{U}_{k,i}^H\bm{y}_{k,i}$.
Then, the received mean square error (MSE) of IU $i$ on SB $k$ is given by
\begin{align}\label{MSE}
\bm{E}_{k,i}& = \mathbb{E}_{\bm{s},\bm{n}} \left[(\hat{\bm{s}}_{k,i} - \bm{s}_{k,i}) (\hat{\bm{s}}_{k,i} - \bm{s}_{k,i})^H\right]  \nonumber \\
&= (\bm{U}_{k,i}^H\bm{Z}_{k,i}\bm{F}_{k,i} - \bm{I}_{d} )^2  + \sum_{u \neq i}^{I} \bm{U}_{k,i}^H\bm{Z}_{k,i}\bm{F}_{k,u}\bm{F}^H_{k,u}\bm{Z}^H_{k,i}\bm{U}_{k,i} + \bm{U}_{k,i}^2\sigma^2_{k,i}.
\end{align}
The optimal MMSE decoding matrix $\{\bm{U}_{k,i}\}$ is given by 
\begin{equation} \label{alpha}
\bm{U}_{k,i} = ( \bm{Z}_{k,i}\bm{F}_{k,i}\bm{F}^H_{k,i}\bm{Z}^H_{k,i} +  \bm{J}_{k,i})^{-1}\bm{Z}_{k,i} \bm{F}_{k,i}.
\end{equation}
Then, substituting (\ref{alpha}) into (\ref{MSE}),  we have  
\begin{equation}
\bm{E}_{k,i}^{\text{min}} =  \bm{I}_{d}-\bm{F}^H_{k,i} \bm{Z}^H_{k,i}( \bm{Z}_{k,i}\bm{F}_{k,i}\bm{F}^H_{k,i}\bm{Z}^H_{k,i} +  \bm{J}_{k,i})^{-1}\bm{Z}_{k,i} \bm{F}_{k,i}.
\end{equation}

According to the relationship between the $\bm{E}_{k,i}^{\text{min}}$ and the SINR shown in \cite{Palomar.2003}, the original Problem (\ref{Prom1}) can be reformulated as Problem (\ref{Prom2}) by introducing a set of auxiliary variables $\{\bm{W}_{k,i}\}$ together with the receiving matrices $\{\bm{U}_{k,i}\}$.
\begin{equation}\label{Prom2}
\begin{split}
	\underset{\bm{\beta}, \bm \phi,\bm{L},\bm{F}_{k,i},\bm{W}_{k,i},\bm{U}_{k,i}}{\text{min} }  
	&\quad   O^{tot} =\sum_{k=1}^K \sum_{i=1}^I \left(\text{tr}\left(\bm{W}_{k,i}\bm{E}_{k,i}\right)-\log|\bm{W}_{k,i}|\right)  \\
	\text{s.t.} & \quad  C1-C4.  
\end{split}
\end{equation}

Although Problem (\ref{Prom2}) has more optimization variables, the objective function of Problem (\ref{Prom2}) is more tractable.
Consequently, Problem (\ref{Prom2}) can be solved by employing the BCD algorithm, where the optimization variables can be iteratively obtained while keeping the others fixed. 
That is to say, we decouple the optimization problem into three subproblems: optimization of the precoding for IUs, RIS's reflecting coefficients and RIS's coordinate.
Note that the receiving matrices $\{\bm{U}_{k,i}\}$ and the auxiliary matrices $\bm{W}_{k,i}$ can be directly solved according to the above analysis. 
Then, the optimal decoding matrix $\{\bm{U}_{k,i}\}$ is given by (\ref{alpha}), and the optimal $\bm{W}_{k,i}^*$ is given by 
\begin{equation}\label{weightW}
\bm{W}^*_{k,i} = (\bm{E}_{k,i}^{\text{min}})^{-1}.
\end{equation}

\subsection{Precoding Matrices Design} \label{PreOpt}

Given the coordinate of RIS $\bm{L}$, RIS's reflecting coefficients, auxiliary matrices $\bm{W}_{k,i}$ and $\{\bm{U}_{k,i}\}$, the precoding matrices are optimized in this section.
By substituting the MSE $\bm{E}_{k,i}$ in (\ref{MSE}) into (\ref{Prom2}) and discarding the constant terms, the precoding matrices $\bm{F}_{k,i}$ are determined by the following problem 
\begin{equation}\label{Prom3}
	\begin{split}
	\underset{\bm{F}_{k,i}}{\text{min} }  
	&\sum_{k=1}^K\sum_{i=1}^I \text{tr}\left(\bm{F}^H_{k,i}\bm{\bar{W}}_{k}\bm{F}_{k,i}\right) - \sum_{k=1}^K\sum_{i=1}^I 2\Re\left[\text{tr}( \bm{\bar{Z}}_{k,i}\bm{F}_{k,i})\right]   \\
	\text{s.t.} & \quad  C2-C4, 
	\end{split}
\end{equation}
where $\bm{\bar{W}}_{k} = \sum_{i=1}^I \bm{Z}^H_{k,i}\bm{U}_{k,i}\bm{W}_{k,i}\bm{U}^H_{k,i}\bm{Z}_{k,i}$ and $\bm{\bar{Z}}_{k,i}  = \bm{W}_{k,i}\bm{U}^H_{k,i}\bm{Z}_{k,i}$.
Although the objective function of Problem (\ref{Prom3})  is convex, the energy harvesting constraints $C3$ and $C4$ are non-convex. 
Note that $\bm{I}-\bm{\Phi}^H_n\bm{\Phi}_n$ is positive definite.
As a result, we adopt the successive convex approximation method by leveraging the first-order Taylor expansions with the given precoding matrix $\bm{\bar{F}}_{k,i}$ as
\begin{align}
 &\text{tr}(\bm{F}^H_{k,i}\bm{B}_k \bm{{F}}_{k,i})   \geq 2 \Re\{  \text{tr}(\bm{\bar{F}}^H_{k,i}\bm{B}_k\bm{F}_{k,i}) \} -  \text{tr}(\bm{\bar{F}}^H_{k,i}\bm{B}_k\bm{\bar{F}}_{k,i}) , \\
& \text{tr}(\bm{{F}}^H_{k,i}\bm{C}_{k,m}\bm{{F}}_{k,i})  \geq 2 \Re\{ \text{tr}( \bm{\bar{F}}^H_{k,i}\bm{C}_{k,m}\bm{F}_{k,i})\} -  \text{tr}(\bm{\bar{F}}^H_{k,i}\bm{C}_{k,m}\bm{\bar{F}}_{k,i}),
\end{align}
where $\bm{B}_k = \eta_k\bm{H}^H_{k}(\bm{I}-\bm{\Phi}^H\bm{\Phi})\bm{H}_{k}$ and $\bm{C}_{k,m} =  \eta_k\bm{Z}^H_{k,m}\bm{Z}_{k,m}$.
Then, $C3$ and $C4$ can be respectively reformulated as 
\begin{align}
&\sum_{k=1}^K \sum_{i =1}^I 2 \Re\{  \text{tr}(\bm{\bar{F}}^H_{k,i}\bm{B}_k\bm{F}_{k,i})\}  \geq \sum_{k=1}^K \sum_{i =1}^I \text{tr}(\bm{\bar{F}}^H_{k,i}\bm{B}_k \bm{\bar{F}}_{k,i})  + P_{RIS}, \label{PwBeam1}\\
&\sum_{k=1}^{K} \sum_{i =1}^I 2 \Re\{ \text{tr}( \bm{\bar{F}}^H_{k,i}\bm{C}_{k,m}\bm{F}_{k,i})\}  \geq  \sum_{k=1}^{K} \sum_{i =1}^I \text{tr}(\bm{\bar{F}}^H_{k,i}\bm{C}_{k,m}\bm{\bar{F}}_{k,i})  +  P^{UE}_m. \label{PwBeam2}
\end{align}
Then, by substituting $C3$ and $C4$ with (\ref{PwBeam1}) and (\ref{PwBeam2}) respectively, Problem (\ref{Prom3}) can be transformed into a series of convex problems, which can be solved by standard tools, such as the CVX.

\subsection{RIS Reflecting Coefficient Optimization}\label{PhasOpt}

Given $\bm{W}_{k,i}$, $\{\bm{U}_{k,i}\}$, $\{\bm{F}_{k,i}\}$ and the RIS's coordinate $\bm{L}$, 
we consider the optimization of RIS's reflecting coefficients $\varphi_n = \beta_n\exp(j\phi_{n})$, where the reflecting matrix is $\bm{\Phi}= \text{diag}\{[\varphi_n ]_{n =1}^N\}$.

According to the MSE given in (\ref{MSE}), we have 
\begin{equation}\label{Weki}
\text{tr}(\bm{W}_{k,i}\bm{E}_{k,i}) =  \text{tr}(\bm{W}_{k,i} \bm{U}_{k,i}^H\bm{Z}_{k,i}\bm{F}^s_{k}\bm{Z}_{k,i}^H\bm{U}_{k,i})  - 2\Re{\left[ \text{tr}(\bm{Z}_{k,i}\bm{F}_{k,i}\bm{W}_{k,i} \bm{U}^H_{k,i} )\right]} + \text{const},
\end{equation}
where $\bm{F}^s_{k} = \sum_{u =1}^{I}\bm{F}_{k,u}\bm{F}^H_{k,u}$. 
The term ``$\text{const}$'' denotes the constant that is irrelevant with the reflecting coefficients $\varphi_n$.
As $\bm{Z}_{k,i} =\bm{H}_{k,i} +\bm{G}_{k,i} $, by removing the irrelevant terms in (\ref{Weki}), the reflecting coefficient optimization problem is formulated as
\begin{equation}\label{Prom8}
	\begin{split}
	\underset{\bm{\varphi}}{\text{min} }  
	&\sum_{k=1}^K\sum_{i=1}^I O_{k,i}(\bm{\varphi}) \\
	\text{s.t.} & \quad  C1,C3,C4,
	\end{split}
\end{equation}
where $\bm{\varphi} = [\varphi_1, \cdots, \varphi_N]^T$.
In the objective function of (\ref{Prom8}), for simplicity, we define $\bm{\bar{U}}_{k,i} = \bm{U}_{k,i}\bm{W}_{k,i}\bm{U}^H_{k,i}$, $\bm{\bar{F}}_{k,i} = \bm{F}_{k,i}\bm{W}_{k,i}\bm{U}_{k,i}^H$, and then we have
\begin{equation}\label{Okiobj}
O_{k,i}(\bm{\varphi}) = 2 \Re[\text{tr}(\bm{G}_{k,i}\bm{F}^s_{k}\bm{H}^H_{k,i}\bm{\bar{U}}_{k,i})]+ \text{tr}(\bm{G}_{k,i}\bm{F}^s_{k}\bm{G}^H_{k,i}\bm{\bar{U}}_{k,i}) 
- 2 \Re[\text{tr}(\bm{G}_{k,i}\bm{\bar{F}}_{k,i})].
\end{equation}
Note that $\bm{\bar{U}}_{k,i} $ is hermitian, but it is still difficult to solve Problem (\ref{Prom8}) with this formulation.

To obtain a more tractable problem formulation, we define $\bm{u}_{k,i} =  \bm{e}_{k,i}^H \odot \bm{e}_{k}^T$.
Then, the RIS assisted channel gain is represented as 
\begin{equation}\label{GkiVec}
\bm{G}_{k,i} = g_{k,i}(\bm{u}_{k,i} \bm{\varphi})\bm{r}_{k,i}\bm{v}_{k}^H.
\end{equation}
Substituting (\ref{GkiVec}) into (\ref{Okiobj}), we have
\begin{equation} \label{Oki}
O_{k,i}(\bm{\varphi}) = 2\Re\{g_{k,i}\xi_{k,i}\bm{u}_{k,i} \bm{\varphi}\} + A_{k,i}(g_{k,i})^2(\bm{u}_{k,i} \bm{\varphi})^2,
\end{equation}
where $A_{k,i} = \text{tr}(\bm{r}_{k,i}\bm{v}_{k}^H\bm{F}^s_{k}\bm{v}_{k}\bm{r}^H_{k,i}\bm{\bar{U}}_{k,i}) $, and 
$\xi_{k,i} = (\text{tr}(\bm{r}_{k,i}\bm{v}_{k}^H\bm{F}^s_{k}\bm{H}^H_{k,i}\bm{\bar{U}}_{k,i}) - \text{tr}(\bm{r}_{k,i}\bm{v}_{k}^H\bm{\bar{F}}_{k,i}) )$.

Similarly, according to (\ref{HRIS}), constraint $C3$ can be reformulated as
\begin{equation}\label{C3S}
\text{tr}\left( \bm{F}^H_{k,i}\bm{H}^H_{k}(\bm{I}-\bm{\Phi}^H\bm{\Phi})\bm{H}_{k}\bm{F}_{k,i} \right)
=  (H_k)^2\text{tr}\left( \bm{F}^H_{k,i} \bm{v}_{k}\bm{v}_{k}^H\bm{F}_{k,i} \right)(N-\bm{\varphi}^H \bm{\varphi}), 
\end{equation}
where $\bm{e}_{k}^H \bm{e}_{k} = N$ is utilized.

Similarly, for constraint $C4$, by substituting $\bm{Z}_{k,m} =\bm{H}_{k,m} +\bm{G}_{k,m}$ and (\ref{GkiVec}) into $C4$,  we have the following reformulation as
\begin{equation}\label{C4S}
\sum_{i=1}^I\text{tr}\left(\bm{Z}_{k,m}\bm{F}_{k,i}\bm{F}^H_{k,i}\bm{Z}^H_{k,m} \right) 
=   (g_{k,m})^2\Lambda_{k,m}(\bm{u}_{k,m} \bm{\varphi})^2  + 2 \Re\{g_{k,m} w_{k,m}\bm{u}_{k,m} \bm{\varphi} \}   +  Q_{k,m} ,
\end{equation}
where $\Lambda_{k,m} =\text{tr}\left( \bm{r}_{k,i}\bm{v}_{k}^H\bm{F}^s_{k}\bm{v}_{k}\bm{r}_{k,i}^H\right)$,
$w_{k,m} = \text{tr}\left(\bm{r}_{k,i}\bm{v}_{k}^H \bm{F}^s_{k} \bm{H}^H_{k,m}\right)$,
and $Q_{k,m} =  \text{tr}\left(\bm{H}_{k,m}\bm{F}^s_{k}\bm{H}^H_{k,m} \right)$.

According to (\ref{GkiVec}), (\ref{C3S}) and (\ref{C4S}), the RIS's reflecting coefficient problem can be reformulated as 
\begin{subequations}\label{Prom7}
	\begin{align}
	\underset{\bm{\varphi}}{\text{min} }  
	&\quad \quad  \bm{\varphi}^H \bm{A} \bm{\varphi}  +  \Re\{\bm{\xi}\bm{\varphi} \} \label{obj7}  \\
	\text{s.t.} & \quad        \bm{\varphi}^H \bm{\Lambda}_m\bm{\varphi} + \Re\{\bm{\omega}_{m} \bm{\varphi}\} \geq \tilde{P}^{U}_m. \label{Prom7_st2}  \\
	& \quad  (N -\bm{\varphi}^H \bm{\varphi})C_{RIS}  \geq  P^{I} \label{Prom7_st1}  \\
	& \quad  |\bm{\varphi}|  \leq \bm{1}.
	\end{align}
\end{subequations} 
where 
\begin{align}
&\bm{A}= \sum_{k=1}^K\sum_{i=1}^I A_{k,i}(g_{k,i})^2\bm{u}^H_{k,i} \bm{u}_{k,i},
\ \bm{\xi} =  \sum_{k=1}^K\sum_{i=1}^I 2g_{k,i}\xi_{k,i}\bm{u}_{k,i},  \nonumber \\
&\bm{\Lambda}_m = \sum_{k=1}^K\eta_k (g_{k,m})^2\Lambda_{k,m}\bm{u}^H_{k,m}\bm{u}_{k,m}, \
\bm{\omega}_{m} = 2\sum_{k=1}^{K} \eta_k g_{k,m}  w_{k,m}\bm{u}_{k,m},\nonumber \\
&\tilde{P}^{U}_m =  {P}^{U}_m  - \sum_{k=1}^{K} Q_{k,m}, \text{and} \ C_{RIS} = \sum_{k=1}^K\sum_{i=1}^I\eta_k (H_k)^2\text{tr}\left( \bm{F}^H_{k,i} \bm{v}_{k}\bm{v}_{k}^H\bm{F}_{k,i} \right). \nonumber
\end{align}
However, it is observed that constraint (\ref{Prom7_st2}) is non-convex.
Note that $\bm{\Lambda}_m$ is positive-definite so that we adopt the first-order Taylor expansion for convex approximation.
At given $\bm{\bar{\varphi}}$, we have 
\begin{equation} \label{phiLine}
\bm{\varphi}^H \bm{\Lambda}_m\bm{\varphi} \geq 2 \Re\{\bm{\varphi}^H \bm{\Lambda}_m \bm{\bar{\varphi}}\} - \bm{\bar{\varphi}}^H \bm{\Lambda}_m\bm{\bar{\varphi}}.
\end{equation}
By utilizing (\ref{phiLine}) to simplify the (\ref{Prom7_st2}), Problem (\ref{Prom7}) can be transformed into a series of simple convex problems, which can be easily solved by CVX. 

\subsection{Optimization of RIS's Coordinate}

We consider the optimization of RIS coordinate with given $\bm{W}_{k,i}$ and $\{\bm{U}_{k,i}\}$, $\{\bm{F}_{k,i}\}$ and the phase shift matrix $\bm{\Phi}$.
In this case, based on the formulations given in (\ref{Oki}), (\ref{C3S}) and (\ref{C4S}), the original Problem (\ref{Prom2}) with respect to the RIS coordinate is formulated as
\begin{subequations}\label{Prom4}
	\begin{align}
	\underset{\bm{L}}{\text{min} }  
	&\quad O^{tot}(\bm{L})=\sum_{k=1}^K\sum_{i=1}^I   {g}_{k,i}(\bm{L})^2E_{k,i}(\bm{L}) + \Re\{{g}_{k,i}(\bm{L})F_{k,i}(\bm{L})\} +  \text{Cst}(\bm{W}_{k,i},\bm{U}_{k,i},\bm{F}_{k,i})    \label{obj4}  \\
	\text{s.t.} & \quad   \sum_{k=1}^K  ({g}_{k,i}(\bm{L})^2 \lambda_{k,m}(\bm{L})  + \Re\{g_{k,m} (\bm{L})\chi_{k,m}(\bm{L})\} + \eta_kQ_{k,m}) \geq P^U_m, \label{Prom4_st2}\\
	& \quad   \sum_{k=1}^K H_{k}(\bm{L})^2 D_k(\bm{L}) \geq P^I,  \label{Prom4_st1}
	\end{align}
\end{subequations} 
where $ \text{Cst}(\bm{W}_{k,i},\bm{U}_{k,i},\bm{F}_{k,i}) $ is the constant term, and  
\begin{align}
& E_{k,i}(\bm{L}) = A_{k,i}(\bm{L})(\bm{u}_{k,i}(\bm{L}) \bm{\varphi})^2,  \label{Eki}  \\
& F_{k,i}(\bm{L}) = 2\xi_{k,i}(\bm{L})\bm{u}_{k,i}(\bm{L}) \bm{\varphi},  \label{Fki}  \\
& \lambda_{k,m}(\bm{L}) = \eta_k \Lambda_{k,m}(\bm{L})(\bm{u}_{k,m}(\bm{L})\bm{\varphi})^2, \label{LambKi} \\
& \chi_{k,m}(\bm{L}) = 2\eta_k w_{k,m}(\bm{L})\bm{u}_{k,m}(\bm{L}) \bm{\varphi}, \label{Chikm} \\
& D_k(\bm{L})  =\sum_{i =1}^I \eta_k\text{tr}\left( \bm{F}^H_{k,i}  \bm{v}_{k}(\bm{L})\bm{v}_{k}^H(\bm{L})\bm{F}_{k,i} \right)(N-\bm{\varphi}^H \bm{\varphi}).\label{Dk} 
\end{align}

According to the channel model, many periodic cosine components with respect to the SB's index and UE's index are involved in $E_{k,i}(\bm{L})$, $F_{k,i}(\bm{L})$, $\lambda_{k,m}(\bm{L})$, $\chi_{k,m}(\bm{L})$ and $D_k(\bm{L})$.
However, they are all dependent on the RIS's coordinate $\bm{L} = [X,Y,Z]$.
Their complex expressions make it very difficult to directly optimize the objective function given in (\ref{obj4}).  
Therefore, we seek to find a tractable formulation of the coordinate optimization problem by regarding these intractable terms as constants.

First of all, for ease of exposition, we define auxiliary variables $r_u$ and $d_0$, which are dependent on the RIS's coordinate $\bm{L}$ by 
\begin{equation}
d_0 = |\bm{L}-\bm{s}_{AP}^1| , r_u = |\bm{L}-\bm{s}_u^1|, u \in \mathcal{I} \cup \mathcal{M}.
\end{equation}
Then, we define the function $f_{k,u}(r_u,d_0)$ with respect to $(r_u,d_0)$ as
\begin{equation} \label{fku}
f_{k,u}(r_u,d_0)= \frac{\mu_k}{ r_u d_0}\exp\left({-K_k(r_u + d_0)}\right), u \in \mathcal{I} \cup \mathcal{M},
\end{equation}
where $\mu_k = \frac{G_tG_r\lambda_k}{8 \sqrt{\pi^3}}$, and $K_k = \frac{K(f_k)}{2}$.

According to (\ref{Gku}), the cascaded channel gain ${g}_{k,i}(\bm{L})$ can be represented as a function with respect to $(r_i,d_0)$ as
\begin{equation}\label{gkid}
{g}_{k,i}(r_i,d_0) =f_{k,i}(r_i,d_0) \exp\left({-j 2\pi \frac{r_i + d_0}{\lambda_k}}\right), u \in \mathcal{I} \cup \mathcal{M}.
\end{equation}

Next, the objective function and constraints are investigated step by step.
As the solution is obtained by the iterative algorithm, we adopt a given RIS's coordinate obtained at the $(n)$-th iteration denoted as $\bm{L}^{(n)}$ to help find the tractable formulation of Problem (\ref{Prom4}).

\subsubsection{Simplification of Objective}
To simplify objective (\ref{obj4}), substituting the coordinate $\bm{L}^{(n)}$ into (\ref{Eki}) and (\ref{Fki}), we can obtain the following \textbf{constants} as
$$E_{k,i} \triangleq E_{k,i}(\bm{L}^{(n)}), F_{k,i} \triangleq \Re\left\{ \exp\left({-j 2\pi \frac{\tilde{r}_i+ \tilde{d}_0}{\lambda_k}}\right) F_{k,i}(\bm{L}^{(n)})\right\}, i \in \mathcal{I},  $$
where the constants $\tilde{r}_i$ and $\tilde{d}_0$ are obtained by leveraging the coordinate $\bm{L}^{(n)}$ in $(n)$-th iteration as
$$\tilde{r}_i =|\bm{L}^{(n)}-\bm{s}_i^1|, \tilde{d}_0 = |\bm{L}^{(n)}-\bm{s}_{AP}^1|.$$

The Hessian matrix of $f_{k,i}(r_i,d_0)$ with respect to $(r_i,d_0)$ is given by
\begin{equation}
\bm{\bigtriangledown}^2f_{k,i} 
= \frac{\mu_k}{r_i d_0}\exp\left({-K_k(r_u + d_0)}\right)\left[ \begin{array}{cc}
\frac{(K_k r_u +1)^2+1}{r_i^2} & \frac{(K_k r_u +1)(K_k d_0 +1)}{r_i d_0}  \\
\frac{(K_k r_u +1)(K_k d_0 +1)}{r_i d_0} & \frac{(K_k d_0 +1)^2+1}{d_0^2} 
\end{array}
\right ].
\end{equation} 
Then, it is observed that $\bm{\bigtriangledown}^2f_{k,i}$ is positive-definite so that $f^2_{k,i}(r_i,d_0)$ is convex with respect to $(r_i,d_0)$.
However, note that the calculated coefficient $F_{k,i}$ is not necessarily positive.
As a result, the first-order Taylor expansion of $f_{k,i}(r_i,d_0)$ is adopted for simplification as
\begin{equation}
f_{k,i}(r_i,d_0) \geq f_{k,i}(\tilde{r}_i,\tilde{d}_0) + \nabla_{r_i} f_{k,i}(\tilde{r}_i, \tilde{d}_0)(r_i - \tilde{r}_i) + \nabla_{d_0} f_{k,i}(\tilde{r}_i, \tilde{d}_0)(d_0 - \tilde{d}_0),  \label{TaylorFF}
\end{equation}
where the derivative is
\begin{equation} \label{DivF1}
\nabla_{x} f_{k,i}(x, y) = -\mu_k\frac{ K_k x +1  }{x^2 y}  \exp\left({-K_k(x + y)}\right).
\end{equation}

In addition, the Hessian matrix of $f^2_{k,i}(r_i,d_0)$ with respect to $(r_i,d_0)$ is 
\begin{equation}
\bm{\bigtriangledown}^2f^2_{k,i} 
= \frac{\mu_k^2}{r^2_i d^2_0}\exp\left({-2K_k(r_u + d_0)}\right)\left[ \begin{array}{cc}
\frac{(2K_k r_u +2)^2+2}{r_i^2} & \frac{(2K_k r_u +2)(2K_k d_0 +2)}{r_i d_0}  \\
\frac{(2K_k r_u +2)(2K_k d_0 +2)}{r_i d_0} & \frac{(2K_k d_0 +2)^2+2}{d_0^2} 
\end{array}
\right ].
\end{equation} 
As $\bm{\bigtriangledown}^2f^2_{k,i}$ is positive definite, $f_{k,i}(r_i,d_0)$ is convex with respect to $(r_i,d_0)$.
Furthermore, it is observed that the calculated coefficient $E_{k,i}$ is always positive. 
Consequently, by leveraging (\ref{gkid}) and (\ref{TaylorFF}), the calculated $E_{k,i}$, $F_{k,i}$ and discarding the irrelevant constants, a simplified version of the objective (\ref{obj4}) with a given coordinate $\bm{L}^{(n)}$ is given as 
\begin{equation} \label{Osd}
O_{\bm{L}^{(n)}}(r_i,d_0) = \sum_{k=1}^K\sum_{i=1}^I (f^2_{k,i}(r_i,d_0)E_{k,i} + F_{k,i}(\nabla_{r_i} f_{k,i}(\tilde{r}_i, \tilde{d}_0)r_i + \nabla_{d_0} f_{k,i}(\tilde{r}_i, \tilde{d}_0)d_0 ) ).
\end{equation} 
It is observed that the objective $O_{\bm{L}^{(n)}}(r_i,d_0)$ is convex with respect to $(r_i,d_0)$.

\subsubsection{Simplification of Constraints for EUs}
Then, we deal with constraint (\ref{Prom4_st2}) for the EU $m$.

By substituting the coordinate $\bm{L}^{(n)}$ into (\ref{LambKi}) and (\ref{Chikm}), the following \textbf{constants}  can be obtained as
$$\lambda_{k,m} \triangleq \lambda_{k,m}(\bm{L}^{(n)}),  \chi_{k,m} \triangleq \Re\left\{ \exp\left({-j 2\pi \frac{\tilde{r}_m+ \tilde{d}_0}{\lambda_k}}\right) \chi_{k,m}(\bm{L}^{(n)}) \right\}, m \in \mathcal{M}, $$
where the constant $\tilde{r}_m =|\bm{L}^{(n)} -\bm{s}_m^1|$ is calculated with the coordinate $\bm{L}^{(n)}$ .

Then, by employing the function $f_{k,m}(r_m,d_0), m \in \mathcal{M}$ in (\ref{fku}), the constants $\lambda_{k,m}$ and $\chi_{k,m}$, constraint (\ref{Prom4_st2}) is rewritten as 
\begin{equation} \label{Convx_P4st2}
 \sum_{k=1}^K ( f^2_{k,m}(r_m,d_0) \lambda_{k,m} + f_{k,m}(r_u,d_0)\chi_{k,m} + \eta_kQ_{k,m}) \geq P^U_m, m \in \mathcal{M}.
\end{equation}
However, (\ref{Convx_P4st2}) is still non-convex.
As $f_{k,m}(r_u,d_0)$ and $f^2_{k,m}(r_u,d_0)$ are both convex functions,  their first-order Taylor expansions are
\begin{align}
f^2_{k,m}(r_m,d_0)&\geq f^2_{k,m}(\tilde{r}_m,\tilde{d}_0) + \nabla_{r_m} f^2_{k,m}(\tilde{r}_m, \tilde{d}_0)(r_m - \tilde{r}_m)  + \nabla_{d_0} f^2_{k,m}(\tilde{r}_m, \tilde{d}_0)(d_0 - \tilde{d}_0),  \label{TaylorF2} \\
f_{k,m}(r_m,d_0) &\geq f_{k,i}(\tilde{r}_m,\tilde{d}_0) + \nabla_{r_m} f_{k,m}(\tilde{r}_i, \tilde{d}_0)(r_m - \tilde{r}_m) + \nabla_{d_0} f_{k,m}(\tilde{r}_m, \tilde{d}_0)(d_0 - \tilde{d}_0),  \label{TaylorFun}
\end{align}
where $\nabla_{x} f_{k,m}(x, y)$ is given in (\ref{DivF1}), and $\nabla_{x} f^2_{k,m}(x, y) =  2 f_{k,m}(x,y)\nabla_{x} f_{k,m}(x, y)$.

Then, substituting (\ref{TaylorF2}) and (\ref{TaylorFun}) into (\ref{Convx_P4st2}), constraint (\ref{Prom4_st2}) for the EU $m$ can be further reformulated as
\begin{align}
A_m(\bm{L}^{(n)},\tilde{r}_m, \tilde{d}_0) r_m + B_m(\bm{L}^{(n)},\tilde{r}_m, \tilde{d}_0) d_0 + C_m(\bm{L}^{(n)},\tilde{r}_m, \tilde{d}_0)  \geq P_m^U, m \in \mathcal{M}, \label{UEPwst}
\end{align}
where
$A_m(\bm{L}^{(n)}, \tilde{r}_m, \tilde{d}_0)  = \sum_{k=1}^K \left(\lambda_{k,m}(\bm{L}^{(n)}) \nabla_{r_m} f^2_{k,m}(\tilde{r}_m, \tilde{d}_0)  + \chi_{k,m}(\bm{L}^{(n)}) \nabla_{r_m} f_{k,m}(\tilde{r}_m, \tilde{d}_0) \right)$, \\
$\quad B_m(\bm{L}^{(n)},\tilde{r}_m, \tilde{d}_0)  = \sum_{k=1}^K \left(\lambda_{k,m}(\bm{L}^{(n)})\nabla_{d_0} f^2_{k,m}(\tilde{r}_m, \tilde{d}_0)  + \chi_{k,m}(\bm{L}^{(n)}) \nabla_{d_0} f_{k,m}(\tilde{r}_m, \tilde{d}_0) \right)$, and \\
$\quad C_m(\bm{L}^{(n)},\tilde{r}_m, \tilde{d}_0)  = \sum_{k=1}^K (\lambda_{k,m}(\bm{L}^{(n)}) f^2_{k,m}(\tilde{r}_m,\tilde{d}_0) +\chi_{k,m}(\bm{L}^{(n)}) f_{k,m}(\tilde{r}_m,\tilde{d}_0) ) - A_m(\bm{L}^{(n)}, \tilde{r}_m, \tilde{d}_0)\tilde{r}_m  - B_m(\bm{L}^{(n)}, \tilde{r}_m, \tilde{d}_0)\tilde{d}_0  +  \sum_{k=1}^K \eta_kQ_{k,m}$.

\subsubsection{Simplification of Constraints for RIS}
Similarly, the constraint (\ref{Prom4_st1}) for the RIS is simplified in the following.
We first define function $h_k(d_0)$ as
\begin{equation}\label{hk}
h_k(d_0)= \frac{\rho_k}{ d_0^2}\exp\left(-2K_kd_0\right),
\end{equation}
where $\rho_k = \left( \frac{\lambda_k}{4 \pi}\right)^2$.
By checking the Hessian matrix of $h_k(d_0)$, it can be verified that $h_k(d_0)$ is convex with respect to $d_0$. 
Also, by substituting the coordinate $\bm{L}^{(n)}$ into (\ref{Dk}), \textbf{constant} $D_k$ can be obtained as $ D_k \triangleq D_k(\bm{L}^{(n)})$.

However, by employing the function $h_k(d_0)$ in (\ref{hk}), and substituting $D_k$ into (\ref{Prom4_st2}), the following constraint is still non-convex:
\begin{equation}
 \sum_{k=1}^K h_k(d_0) D_k \geq P^I .
\end{equation}

Therefore, the first-order Taylor expansion of $h_k(d_0)$ is utilized for convex approximation, which is
\begin{align}
h_k(d_0)&\geq h_k(\tilde{d}_0) + \nabla_{d_0}^{h_k}(\tilde{d}_0)(d_0 - \tilde{d}_0), \label{TaylorH} 
\end{align}
where $\nabla_{d_0}^{h_k}(\tilde{d}_0)$ represents the first-order derivative of $h_k(d_0)$ given by
$$\nabla_{d_0}^{h_k}(d_0) = -\rho_k \frac{2K_kd_0 +2}{ d_0^3}\exp\left(-2K_kd_0\right).$$
Then, substituting (\ref{TaylorH}) into (\ref{Prom4_st1}), we have
\begin{align}
A_{RIS}(\bm{L}^{(n)},\tilde{d}_0)d_0 + B_{RIS}(\bm{L}^{(n)},\tilde{d}_0)\geq P^I,   \label{RISPwst}
\end{align}
where the constants are evaluated as
\begin{align}
&A_{RIS}(\bm{L}^{(n)},\tilde{d}_0) = \sum_{k=1}^K D_k(\bm{L}^{(n)}) \nabla_{d_0}^{h_k}(\tilde{d}_0),  \label{AIRS}\\
&B_{RIS}(\bm{L}^{(n)},\tilde{d}_0) = \sum_{k=1}^K D_k(\bm{L}^{(n)}) ( h_k(\tilde{d}_0) - \tilde{d}_0 \nabla_{d_0}^{h_k}(\tilde{d}_0)).
\end{align}

\subsubsection{Simplification of Problem (\ref{Prom4})}

Finally, given RIS's coordinate obtained at the $(n)$-th iteration denoted as $\bm{L}^{(n)}$, by replacing  $O^{tot}(\bm{L})$ with $O_{\bm{L}^{(n)}}(r_i,d_0)$ in (\ref{Osd}), replacing (\ref{Prom4_st2}) and (\ref{Prom4_st1}) 
with (\ref{UEPwst}) and (\ref{RISPwst}) respectively, Problem (\ref{Prom4})  is reformulated as
\begin{subequations}\label{Prom5}
	\begin{align}
	\underset{\bm{L},r_u,d_0}{\text{min} }  
	&\quad \quad   O_{\bm{L}^{(n)}}(r_i,d_0) \label{obj5}  \\
	\text{s.t.} & \quad |\bm{L}-\bm{s}_u^1|\leq r_u, u \in\mathcal{I}\cup \mathcal{M}  \label{StDu} \\
	 &  \quad|\bm{L}-\bm{s}_{AP}^1|\leq d_0.  \label{StD0}\\
	 & \quad (\ref{UEPwst}),(\ref{RISPwst}).  \nonumber
	\end{align}
\end{subequations} 
Constraints (\ref{StDu}) and (\ref{StD0}) are introduced for the auxiliary variables $r_u$ and $d_0$.
Then, it can be verified that Problem (\ref{Prom5}) is convex, which can be readily solved by CVX.
 
However, note that Problem (\ref{Prom5}) is the simplified version of Problem (\ref{Prom4}).
The optimal solution to Problem (\ref{Prom5}) denoted by $(\bm{L}^*,r_u^*,d_0^*)$ may not satisfy all the constraints of Problem (\ref{Prom4}). 
Consequently, we need to add the following procedure to ensure that the obtained solution $(\bm{L}^*,r_u^*,d_0^*)$ is a feasible solution to Problem (\ref{Prom4}).

\subsubsection{Feasibility guarantee}
First we define the following penalty indicators as
\begin{align}
&\alpha = A_{RIS}(\bm{L}^*,d^*_0) d^*_0 + B_{RIS}(\bm{L}^*,d^*_0) -P^I, \label{a}\\
&\alpha_m = A_m(\bm{L}^*,r^*_m, d^*_0)r^*_m + B_m(\bm{L}^*,r^*_m, d^*_0)d^*_0 + C_m(\bm{L}^*,r^*_m, d^*_0) - P_m^U.\label{am}
\end{align}

If $\alpha < 0$, the obtained coordinate $\bm{L}^*$ does not satisfy constraint (\ref{Prom4_st1}).
This implies that the distance between RIS and AP is too large, i.e., the power harvested by the RIS does not exceed the requirement.
By checking the first-order derivative of $h_k(d_0)$ with respect to $d_0$ , it is verified that $\nabla_{d_0}^{h_k}( d) < 0, \forall d \geq 0$.
According to (\ref{AIRS}), as $D_k(\bm{L}^*) \geq 0$ for all coordinates, so that $A_{RIS}(\bm{L}^*,d^*_0) < 0$.
Consequently, to ensure that (\ref{Prom4_st1}) is satisfied, the RIS should be closer to the AP, i.e., $d_0$ should be reduced.
To this end, the required harvested power of RIS, i.e., $P^I$ is modified as  
\begin{equation}\label{ARIS}
{P^I}' = P^I + \epsilon, \text{if  }\alpha  < 0,
\end{equation}
where $\epsilon >0$ is the introduced penalty. \footnote{In practise, penalties are set as $1\%$ of the required powers for satisfactory performance, i.e., $\epsilon = 1\%P^I$ and $\epsilon_m = 1\%P_m^U$.  }

Also, if $\alpha_m < 0$, the obtained coordinate $\bm{L}^*$ does not satisfy constraint (\ref{Prom4_st2}) for EU $m$.
Note that $ \nabla_{x} f_{k,m}(x, y) < 0$, $ \nabla_{y} f_{k,m}(x, y) < 0$ and $\lambda_{k,m}(\bm{L})\geq 0,  \forall \bm{L}$.
Similarly, according to (\ref{UEPwst}), this implies that $d_0$ or $r_m$ should be decreased to satisfy constraint (\ref{Prom4_st2}).
Given that $d_0$ is modified by adjusting $P^I$, therefore, we only adjust the $P_m^U$ to reduce $d_m$, which should be modified as 
\begin{equation}\label{Am} 
{P_m^U}' = P_m^U + \epsilon_m, \text{if  } \alpha_m  < 0,
\end{equation}
where $\epsilon_m >0$ is the penalty for EU $m$.

Overall, constraints (\ref{RISPwst}) and (\ref{UEPwst}) are updated by replacing ${P^I}$ and ${P_m^U}$ with ${P^I}'$ and ${P_m^U}'$ in (\ref{ARIS}) and (\ref{Am}), respectively.
The solution to Problem (\ref{Prom5}) with the modified constraints  (\ref{RISPwst}) and (\ref{UEPwst}) should be updated accordingly.
Then, the finally obtained solution $(\bm{L}^*,r_u^*,d_0^*)$ is guaranteed to satisfy all the constraints of Problem (\ref{Prom4}) when $\alpha \geq 0$ and $\alpha_m \geq 0, \forall m$.

\emph{Remark} 1: 
Once the obtained $(\bm{L}^*,r_u^*,d_0^*)$ is a feasible solution to the original Problem (\ref{Prom4}). We denote it as the $(n+1)$-th coordinate of the RIS, i.e., $\bm{L}^{(n+1)} =\bm{L}^*$ when $\alpha \geq 0$ and $\alpha_m  \geq 0, \forall m$.
Then, Problem (\ref{Prom5}) can be formulated based on $\bm{L}^{(n+1)}$, and the feasible coordinate of RIS for the $(n+2)$-th iteration can be obtained by leveraging the penalties for power harvesting constraints.

\emph{Remark} 2: In addition, as the objective function $O_{\bm{L}^{(n)}}(r_i,d_0)$ is the approximation of (\ref{obj4}), we need to check the original objective value $O^{tot}(\bm{L}^{(n+1)})$ in (\ref{obj4}) at each iteration.
Note that the receiving matrix $\bm{U}^{n}_{k,i}$ aims to minimize the MSE for a given channel $\bm{Z}(\bm{L}^{(n)})$, and weight $\bm{W}^{(n)}_{k,i}$ is dependent on the current MMSE matrix.
As the channel is optimized by adopting the RIS's new coordinate $\bm{L}^{(n+1)}$, the receiving matrix $\bm{U}_{k,i}$ matched to this new channel and the obtained MMSE should be updated according to (\ref{alpha}) and (\ref{weightW}), respectively. 
That is to say, the objective value at the $(n+1)$-th iteration is evaluated as $O^{tot}(\bm{L}^{(n+1)},\bm{W}^{n+1}_{k,i},\bm{U}^{n+1}_{k,i})$.

\emph{Remark} 3: 
Denote the initial coordinate of RIS as $\bm{L}^{(0)}$.
To guarantee the monotonicity of RIS's position optimization, we need to find the optimized coordinate $\bm{L}^{(n')}$ which can reduce the objective $O^{tot}$ compared with the initial coordinate $\bm{L}^{(0)}$. 
That is to say\footnote{As $O^{tot}$ consists of two parts with different physical meanings: weighted MSE and rate, their numerical values may have more than 3 orders of magnitude difference. In this case, it is better to check these two parts separately for satisfactory performance. }, 
\begin{equation}\label{ObjCmp}
O^{tot}(\bm{L}^{(n')},\bm{W}^{n'}_{k,i},\bm{U}^{n'}_{k,i}) < O^{tot}(\bm{L}^{(0)},\bm{W}^{0}_{k,i},\bm{U}^{0}_{k,i}) .
\end{equation}
Although the obtained objective in $(n+1)$-th iteration may be larger than that of the $(n)$-th iteration, we still update Problem (\ref{Prom5}) by utilizing $\bm{L}^{(n+1)}$ for the next feasible solution when the coordinate $\bm{L}^{(n')}$ satisfying (\ref{ObjCmp}) have not be found.

\emph{Remark} 4: 
Note that Problem (\ref{Prom5}) may become infeasible due to the penalties added in constraints (\ref{RISPwst}) and (\ref{UEPwst}).   
Consequently, it may exist the case that the coordinate $\bm{L}^{(n')}$ satisfying (\ref{ObjCmp}) cannot be found.
In this case, the RIS should keep the initial coordinate $\bm{L}^{(0)}$, i.e.,  $\bm{L}^{(n')}= \bm{L}^{(0)}$.

Overall, the above analysis can be summarized as the following Penalty Constrained Convex Approximation (PCCA) Algorithm \ref{alg1} to optimize the RIS's coordinate.

\begin{algorithm}
	\caption{Penalty Constrained Convex Approximation Algorithm (PCCA)}
	\begin{algorithmic}[1]\label{alg1}
		\STATE Initialize  coordinate $\bm{L}^{(0)}$, and the objective $O^{tot}(\bm{L}^{(0)},\bm{W}^{0}_{k,i},\bm{U}^{0}_{k,i})$;
		\STATE Initialize iterative number $n=0$ and maximum number of iterations $N_{max}$; 
		\REPEAT  
			\STATE Obtain $(\bm{L}^*,r_u^*,d_0^*)$ by solving Problem (\ref{Prom5});
			\IF{$\alpha\geq 0$ and $\alpha_m \geq 0$}  
				\STATE Update $\bm{L}^{n+1} =\bm{L}^*$;
				\STATE Calculate $\bm{W}^{n+1}_{k,i},\bm{U}^{n+1}_{k,i}$ according to  (\ref{alpha}) and (\ref{weightW}), respectively.  
				\STATE Calculate $O^{tot}(\bm{L}^{n+1},\bm{W}^{n+1}_{k,i},\bm{U}^{n+1}_{k,i})$; 
				\IF{$O^{tot}(\bm{L}^{n+1},\bm{W}^{n+1}_{k,i},\bm{U}^{n+1}_{k,i}) < O^{tot}(\bm{L}^{(0)},\bm{W}^{0}_{k,i},\bm{U}^{0}_{k,i})$} 
					\STATE $\bm{L}^{(n')} = \bm{L}^{n+1}$;
				\ENDIF
			\ELSE  
				\STATE Update $P_m^U$ for all $m, \alpha_m <0 $ according to (\ref{am}) ;
				\STATE Update $P^I$ according to (\ref{a}) if $\alpha <0$;
				\IF{ Problem (\ref{Prom5}) is not feasible}
					\STATE Set $n = N_{max}$ and $\bm{L}^{(n')} =\bm{L}^{0}$;
				\ENDIF
			\ENDIF	
		\UNTIL $n = N_{max}$;
		\ENSURE $\bm{L}^{(n')}$,$\bm{U}^{n'}_{k,i},\bm{W}^{n'}_{k,i}$;
	\end{algorithmic}
\end{algorithm}

\subsection{BCD Algorithm to Solve Problem (\ref{Prom2})}

Based on the above analysis, a BCD based alternating optimization algorithm is proposed for alternately optimizing the precoding matrices of the AP, the phase shifts of the RIS and the coordinate of RIS.
The detailed algorithm is presented as in Algorithm \ref{alg2}.

\begin{algorithm}
\caption{The BCD Algorithm to Solve Problem (\ref{Prom2})}
\begin{algorithmic}[1]\label{alg2}
\STATE Initialize feasible $\bm{L}^0$, $\bm{\varphi}^0$, and $\bm{F}^0_{k,i}$.
\STATE Initialize $\bm{U}^{(0)}_{k,i}$ and $\bm{W}^{(0)}_{k,i}$ according to (\ref{alpha}) and (\ref{weightW}), respectively.
\STATE Initialize maximum number of iterations $S_{max}$ and the iterative number $s=0$.
\REPEAT
\STATE Calculate $\bm{F}_{k,i}^{(s+1)}$ by solving the convex approximations of Problem (\ref{Prom3}); 
\STATE Calculate $\bm{\varphi}^{(s+1)}$ by solving the convex approximations of Problem (\ref{Prom7}); 
\STATE Calculate $\{\bm{L}^{(s+1)},\bm{U}^{(s+1)}_{k,i},\bm{W}^{(s+1)}_{k,i}\}$ according to the PCCA Algorithm;
\UNTIL  $s = S_{max}$.
\end{algorithmic}
\end{algorithm}

 The proposed BCD algorithm is guaranteed to converge. 
 Specifically, the generated solutions $\{\bm{L}^{(s)}, r^{(s)}_u, d^{(s)}_0\}$ by the PCCA algorithm are always feasible to the coordinate optimization problem (\ref{Prom4}) by updating $P_m^U$ and $P^I$.
 Furthermore, according to the steps 15-17 and steps 9-11 of the PCCA algorithm, the obtained objective value $\{O^{tot}(\bm{L}^{s},\bm{W}^{s}_{k,i},\bm{U}^{s}_{k,i})\}$ is monotonically decreasing.
 Then, according to Section \ref{PreOpt} and Section \ref{PhasOpt}, Problem (\ref{Prom3}) and Problem (\ref{Prom7}) are reformulated based on the Taylor-expansion based convex optimization.
It can be readily verified that the sequence of solutions  generated by the BCD Algorithm is always feasible for Problem (\ref{Prom2}). 
 The monotonic property of the BCD Algorithm can be similarly proved by using the method in \cite{Pan.2017}.

\section{Simulation Results}	

Simulation results are presented in this section to evaluate the performance of  the proposed algorithm.
In the simulation, THz frequency range at the AP is $300$-$340$ GHz, the bandwidth of each sub-band is $20$ GHz, and the molecular absorption coefficients are generated according to \cite{Boulogeorgos.20186252018628}.
The AP is located along the Y-axis with height of $2$ m.
The AP's transmit antenna is modelled as a uniform planar array (UPA) with size of $N_t = 5 \times 10$, and $P_t^{max} = 10$ W.  
The IUs and EUs are both equipped with $2$ receive antennas.
The antenna gain is set as $G_t = 15$ and $G_r =6$.
The separations between the transmit/receive antennas are set to be $0.1$ mm.

\begin{figure}
		\centering
		\includegraphics[width=0.6\textwidth]{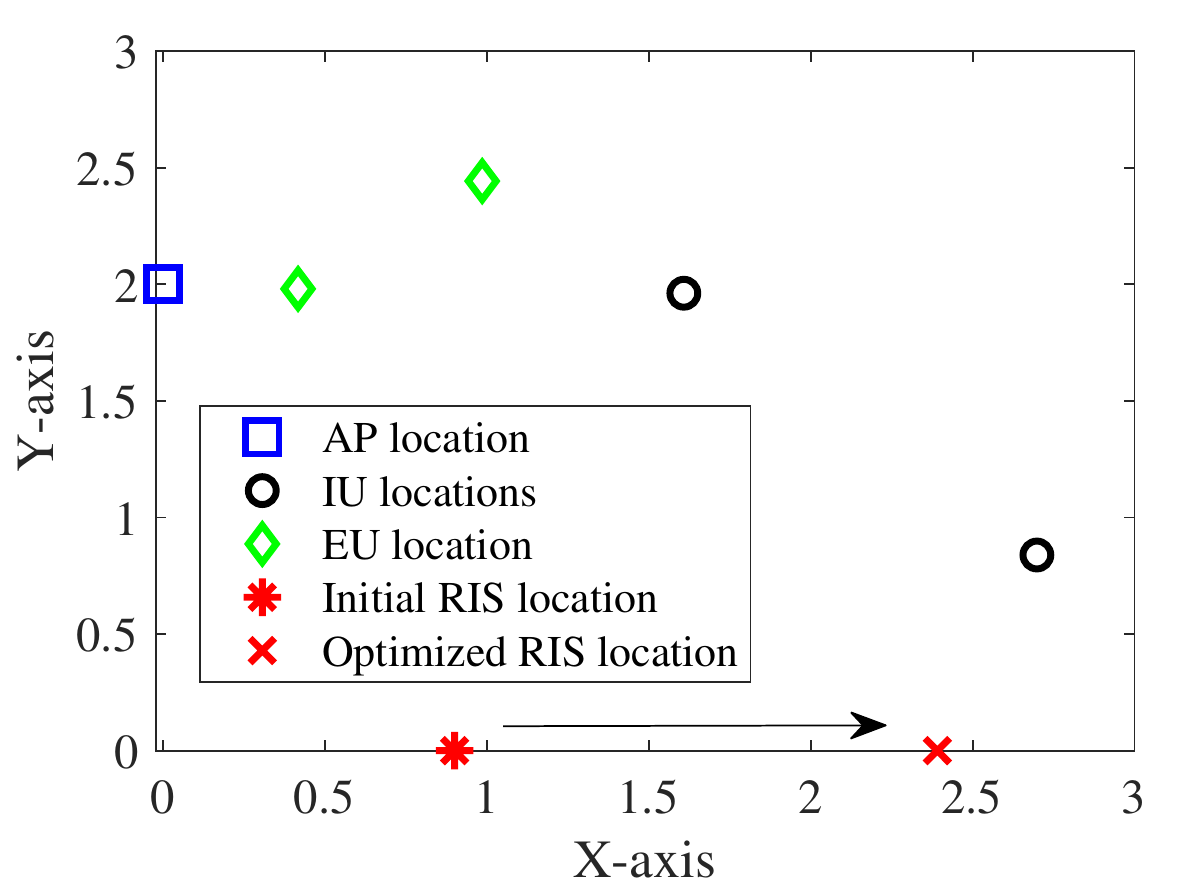}
		\vspace{-1em}
		\caption{{The simulation scenario of the STIPT system.}}
		\vspace{-1em}
		\label{fig2}
	\end{figure}

As shown in Fig. \ref{fig2}, there are $2$ IUs and $2$ EUs which are randomly distributed in a square area with width $3$ m.
The RIS is installed on the X-axis, and the separations between the RIS reflecting elements are $0.1$ mm.
In Fig. \ref{fig2}, the number of reflecting elements is set to $100$.
The initial coordinate of the RIS in X-axis is marked by ``\textasteriskcentered'' and the optimized coordinate in X-axis is marked by ``$\times$''.
It is observed that in the layout shown in Fig. \ref{fig2}, the optimized coordinate of the RIS is updated by the proposed BCD algorithm.

The proposed BCD algorithm given in Algorithm \ref{alg2} is labelled as ``PropBCD''.
For performance comparison, we consider two benchmark schemes:
\begin{itemize}
\item The first scheme which is labelled as ``BeamOpt'' only optimizes the transmit precoding matrices with the fixed RIS's phase shift and coordinate; This scheme can be obtained by removing step 6 and step 7 of Algorithm \ref{alg2}.
\item The scheme labelled as ``FixedLoc'' optimizes both the transmit precoding matrices and the RIS's phase shift, where the RIS's coordinate is kept fixed.
This scheme is obtained by removing step 7 of Algorithm \ref{alg2}.
\end{itemize}

	\begin{figure}
		\centering
		\includegraphics[width=0.6\textwidth]{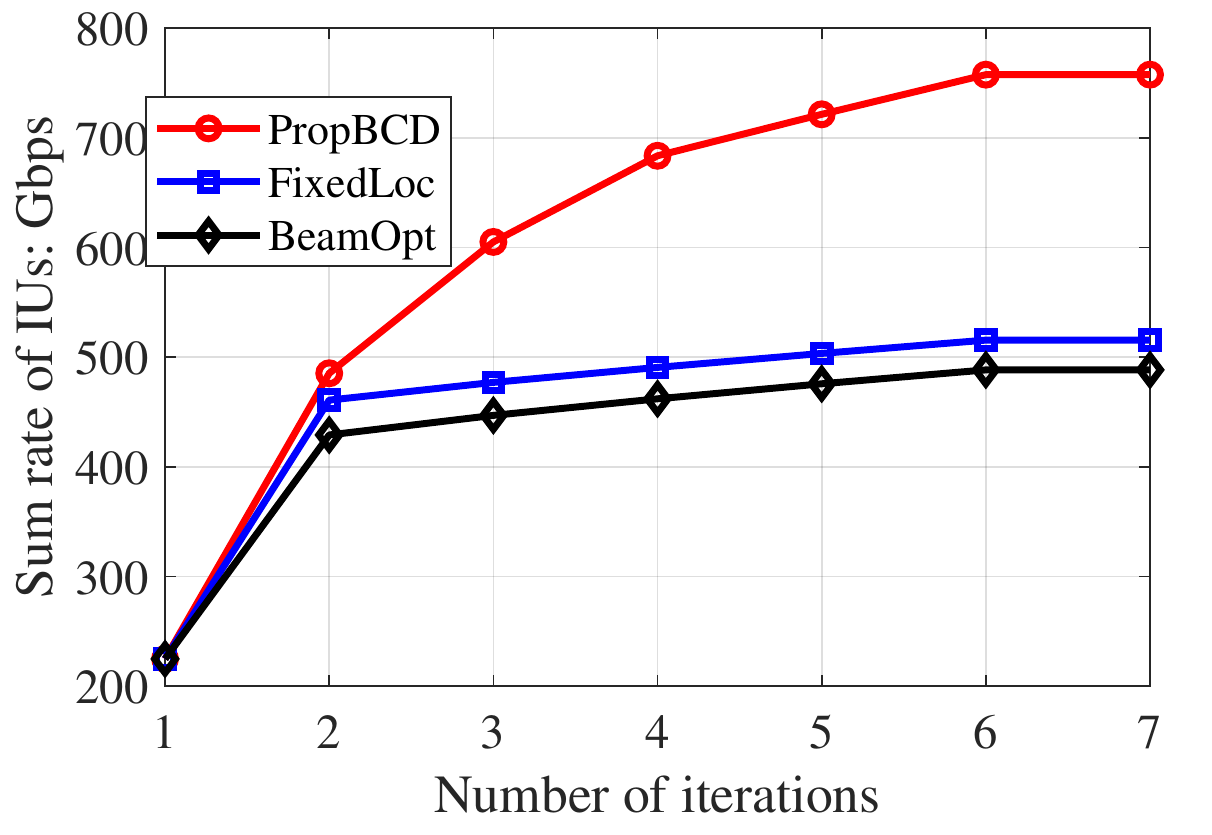}
		\vspace{-1em}
		\caption{The convergence performance of different algorithms.}
		\vspace{-1em}
		\label{fig3}
	\end{figure}

Fig. \ref{fig3} shows the convergence performance of the proposed algorithm and the benchmarks, where its simulation scenario is shown in Fig. \ref{fig2}.
In Fig. \ref{fig3}, the required harvest power by RIS and the EU are set as $0.1$ mW.
It is observed that the achieved sum rate of IUs increases with the number of iterations for all considered cases, and all the considered schemes converge within 7 iterations.
As expected, the proposed BCD algorithm achieves the best performance.
Significant rate improvement can be obtained by optimizing the RIS's coordinate.

	\begin{figure}
		\centering
		\includegraphics[width=0.6\textwidth]{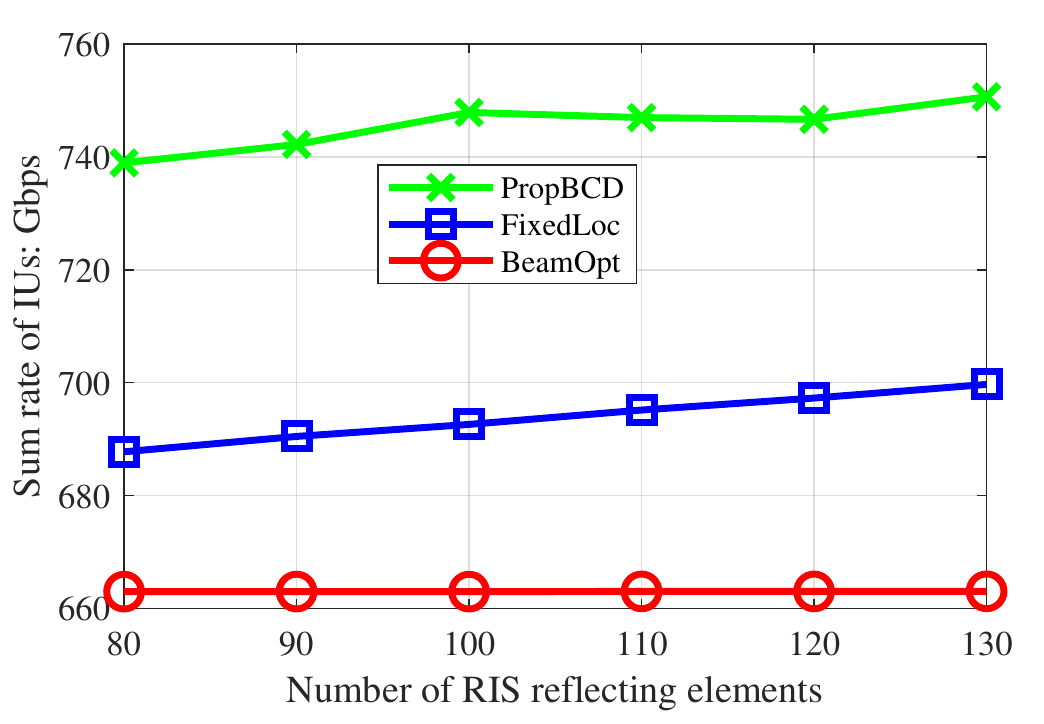}
		\vspace{-1em}
		\caption{The sum rate versus the number of reflecting elements.}
		\vspace{-1em}
		\label{fig4}
	\end{figure}

	\begin{figure}
		\centering
		\includegraphics[width=0.6\textwidth]{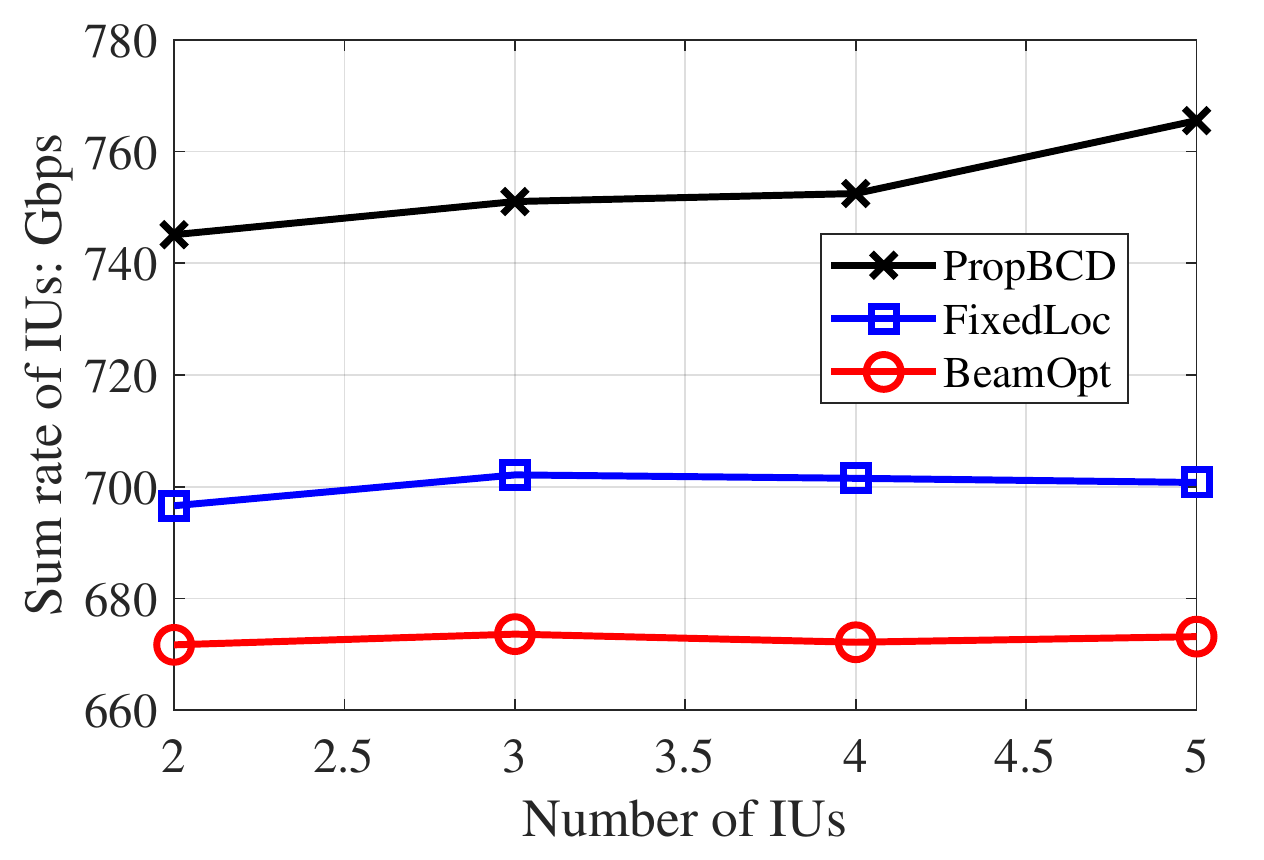}
		\vspace{-1em}
		\quad\caption{The sum rate versus the number of IUs.}
		\vspace{-1em}
		\label{fig5}
	\end{figure}

Fig. \ref{fig4} shows the achieved sum rates of IUs by different schemes versus the number of RIS reflecting elements.
The simulation results are averaged by 100 random realizations, where the initial coordinates of the RIS, the IUs, and the EUs are randomly generated.
As shown in Fig. \ref{fig4},  the sum rates of IUs achieved by the proposed BCD algorithm and the ``FixedLoc'' algorithm increase with the number of reflecting elements.
However, the proposed BCD algorithm can achieve a higher sum rate, and the performance gap increases with the number of reflecting elements. 
This implies that the proposed BCD algorithm can fully exploit the potential benefits provided by the RIS, especially in this STIPT system.
In addition, it is observed that the sum rate of the ``BeamOpt'' algorithm remains the same as the number of reflecting elements increases since the RIS related parameters are not optimized in this scheme.

Fig. \ref{fig5} shows the achieved sum rates of the IUs by different schemas versus the number of IUs in the system.
It is interesting to see that the achieved sum rates by  the ``PropBCD'' algorithm increases with the number of IUs.
Meanwhile the sum rate obtained by the ``BeamOpt'' algorithm slightly increases, and that of the ``FixedLoc'' algorithm keeps fixed.
This result clearly validates the benefits provided by the RIS in the STIPT system.
By utilizing the RIS, the transmission channel can also be optimized to fully exploit the spatial diversity so that the sum rate performance can be enhanced. 
In particular, in the THz system, where the channel gain is very sensitive to the transmit distance, optimizing the RIS's coordinate can help provide considerable performance gain.

\begin{figure}
	\centering
	\includegraphics[width=0.6\textwidth]{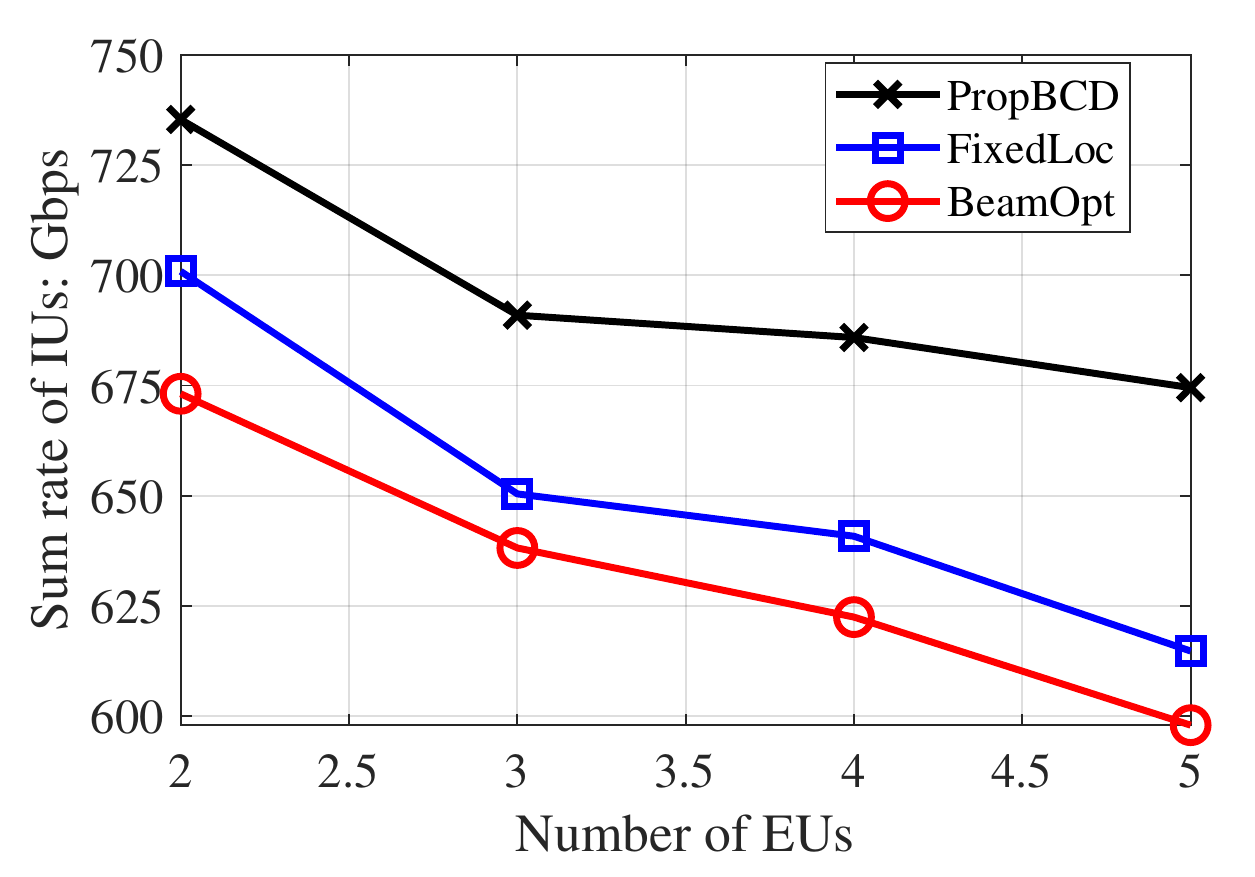}
	\vspace{-1em}
	\caption{The sum rate versus the harvested power of EUs.}
	\vspace{-1em}
	\label{fig55}
\end{figure}

Fig. \ref{fig55} shows the achieved sum rates of the IUs versus the number of EUs in the system, and the number of IUs is fixed to be 2.
Then, it is observed that the sum rate of the IUs decreases with the number of EUs for all cases, as more power needs to be harvested for the EUs.
In addition, the performance gap between the ``PropBCD'' algorithm and the other benchmarks increases with the number of EUs, which also shows the superiority of proposed algorithm.

	\begin{figure}
		\centering
		\includegraphics[width=0.6\textwidth]{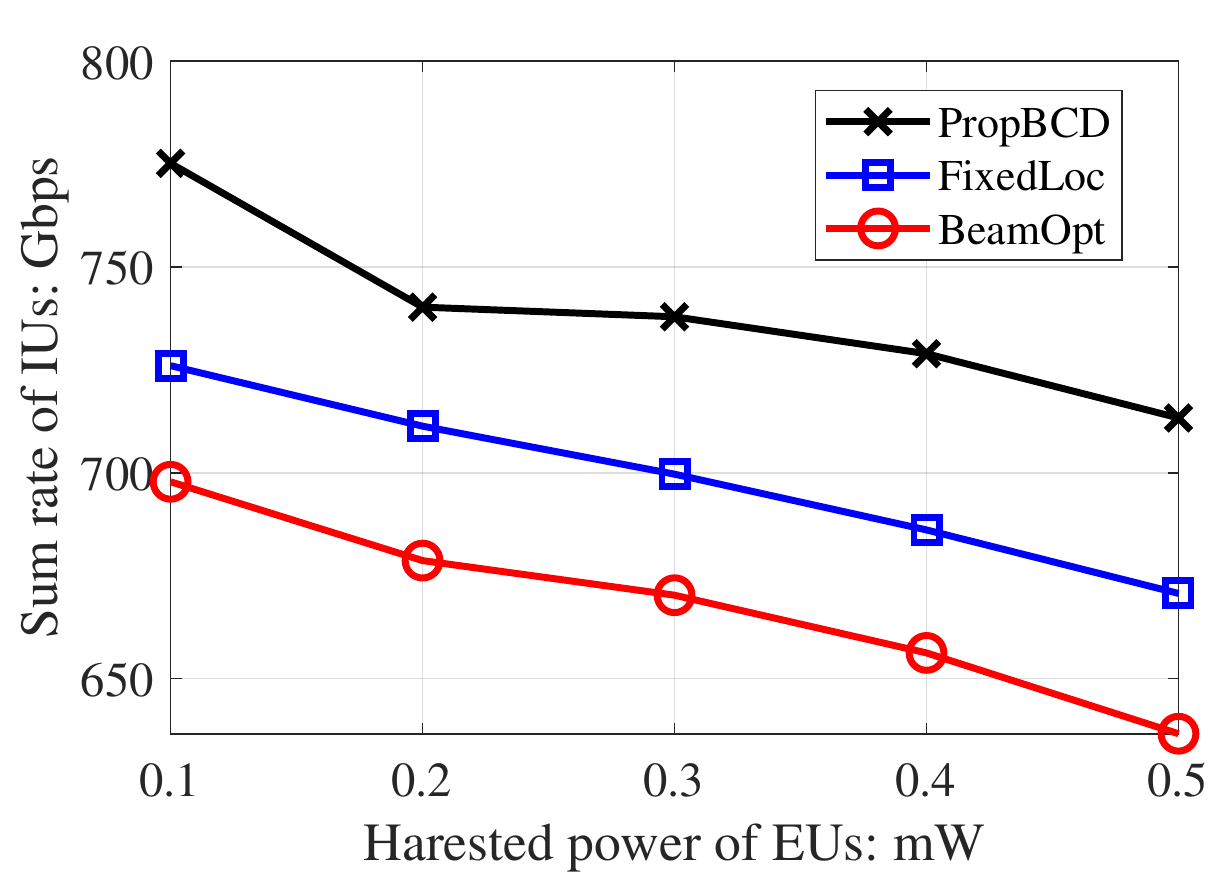}
		\vspace{-1em}
		\caption{The sum rate versus the harvested power of EUs.}
		\vspace{-1em}
		\label{fig6}
	\end{figure}

	\begin{figure}
		\centering
		\includegraphics[width=0.6\textwidth]{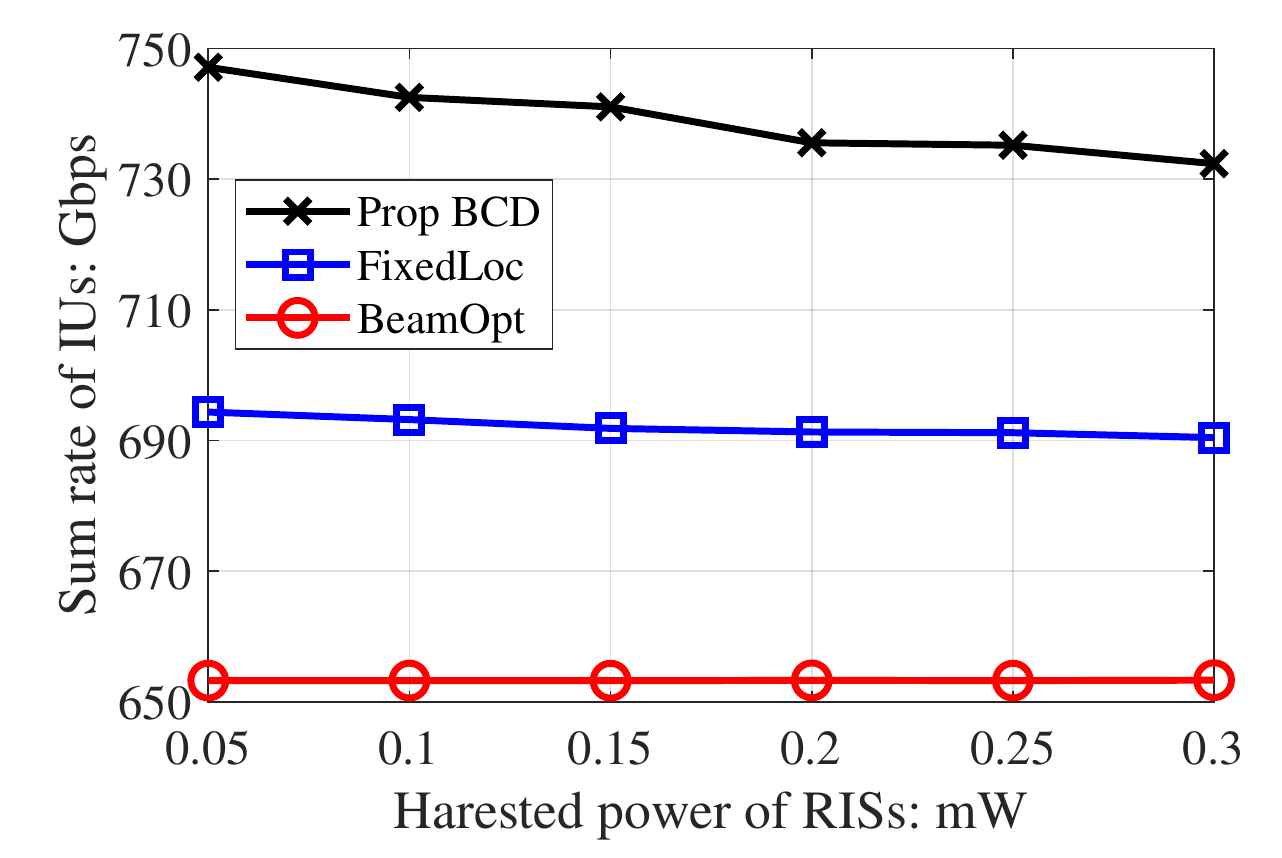}
		\vspace{-1em}
		\caption{The sum rate versus the harvested power of RISs}
		\vspace{-1em}
		\label{fig7}
	\end{figure}

Fig. \ref{fig6} shows the impact of the harvest power required by the EUs on the sum rate performance.
In Fig. \ref{fig6}, the required power of the RIS is fixed to $0.1$ mW, and the number of EUs is $2$.
It is observed that the proposed BCD algorithm outperforms the other two algorithms.
As shown in Fig. \ref{fig6}, the achieved sum rate of IUs by the BCD algorithm decreases with the required harvested power of EUs for all considered cases.
In addition, the performance gap between the ``PropBCD'' algorithm and the other schemes keeps stable with the required power.
This implies that the proposed algorithm still has stable performance advantages when the power harvesting demand from EUs increases.

Finally, Fig. \ref{fig7} shows the impact of the harvest power required by the RIS on sum rate performance.
In Fig. \ref{fig7}, the required power of the EU is $0.1$ mW, and the other simulation parameters are the same as those of Fig. \ref{fig6}.
As expected, the proposed BCD algorithm outperforms the other two algorithms, but the performance gap slightly decreases with the required power to harvest.
It becomes more difficult to find the new coordinate of RIS to improved IU's rate while guaranteeing the more stringent harvested power constraints of RIS. 
Also, compared with the impact of the EU's harvested power shown in Fig. \ref{fig6}, RIS's required harvest power has a slighter impact on rate performance.

\section{Conclusions}

In this paper, we have investigated a new simultaneous THz information and power transfer system, named as STIPT, where the RIS is utilized to support the THz transmission.
In this system, the RIS can utilize the power harvesting technology to self-sustain its power consumption.
The optimization problem has been formulated to maximize the IUs' sum rate while guaranteeing the EU's and RIS's power harvesting requirements.
A BCD-based alternating optimization algorithm has been proposed to optimize the transmit precoding for IUs, the RIS's reflecting coefficients and the RIS's coordinate.
Simulation results have shown that the proposed algorithm can achieve considerable performance gain in terms of the sum rate. 
With the assistance of RIS, the transmission channel can be optimized to fully exploit the spatial diversity so that the sum rate performance can be enhanced. 
As the channel gain of the THz transmission is very sensitive to the transmission distance, optimizing the RIS's coordinate in the STIPT system can help provide considerable performance gain.

\bibliographystyle{ieeetran}
\bibliography{Reference}
	
\end{document}